 \documentclass[10pt,preprint]{aastex}  %compact preprint style (MJA)
 \usepackage{natbib}
\def\etal{et al.~}

\def\gapprox{\lower.4ex\hbox{$\;\buildrel >\over{\scriptstyle\sim}\;$}}
\def\lapprox{\lower.4ex\hbox{$\;\buildrel <\over{\scriptstyle\sim}\;$}}

\shortauthors{ASCHWANDEN}
\shorttitle{Thresholded Power laws}

\begin{document}
%{\sl  Manuscript, accepted ... }

\title{       	Thresholded Power Law Size Distributions of Instabilities 
		in Astrophysics	}

\author{        Markus J. Aschwanden$^1$} 

\affil{		$^1)$ Lockheed Martin, 
		Solar and Astrophysics Laboratory, 
                Org. A021S, Bldg.~252, 3251 Hanover St.,
                Palo Alto, CA 94304, USA;
                e-mail: aschwanden@lmsal.com }

\begin{abstract}
Power law-like size distributions are ubiquitous in astrophysical 
instabilities.  There are at least four natural effects that cause 
deviations from ideal power law size distributions, which we model here 
in a generalized way: (1) a physical threshold of an instability; (2) 
incomplete sampling of the smallest events below a threshold $x_0$; 
(3) contamination by an event-unrelated background $x_b$; and 
(4) truncation effects at the largest events due to a finite system size.
These effects can be modeled in simplest terms with a 
``thresholded power law'' distribution function (also called generalized 
Pareto [type II] or Lomax distribution), $N(x) dx \propto (x+x_0)^{-a} dx$, 
where $x_0 > 0$ is positive for a threshold effect, while $x_0 < 0$ is 
negative for background contamination. 
We analytically derive the functional shape of this thresholded power law 
distribution function from an exponential-growth evolution model, which
produces avalanches only when a disturbance exceeds a critical 
threshold $x_0$. We apply the thresholded power law distribution function 
to terrestrial, solar (HXRBS, BATSE, RHESSI), and stellar flare (Kepler)
data sets. We find that the thresholded power law model provides
an adequate fit to most of the observed data. Major advantages of
this model are the automated choice of the power law fitting range, 
diagnostics of background contamination, physical inastability thresholds, 
instrumental detection thresholds, and finite system size limits. When testing 
self-organized criticality models, which predict ideal power laws, we 
suggest to include these natural truncation effects.
\end{abstract}


\section{		INTRODUCTION				}

Power law-like size distributions of extreme events have been 
discovered over the last 30 years in a large number of astrophysical 
phenomena in many wavelengths, such as hard X-ray and gamma-ray bursts 
from solar and stellar flares, auroral emission and geomagnetic 
substorms from planetary magnetospheres, particle events from 
radiation belts, pulsar glitches, soft gamma-ray repeaters, blazars, 
black-hole objects, cosmic rays, to boson clouds (for a recent review
see Aschwanden et al.~2015 and references therein). 

A unifying concept to explain and model these observed power law 
distributions is the concept of {\sl self-organized criticality (SOC)} 
in slowly-driven nonlinear dissipative systems, originally proposed by 
Bak, Tang and Wiesenfeld (1987; BTW) and Katz (1986). The SOC concept is a very 
interdisciplinary subject, being applied in 
astrophysics (Aschwanden et al.~2015), 
geophysics (Turcotte 1999),
as well as in financial physics (Feigenbaum 2003), 
and social sciences (Galam 2012).
On the most general level,
the common denominator of all SOC processes in different science disciplines
is the statistics of scale-free nonlinear processes, which can be characterized 
by a nonlinear growth phase, during which coherent growth is enabled and 
has multiplicative characteristics, in contrast to linear processes with
incoherent and additive characteristics. Incoherent random
processes exhibit binomial, Gaussian, Poissonian, or exponential
distribution functions, while coherent processes exhibit scale-free
power law-like distributions. This is the fundamental property that earthquakes, 
solar flares, or stock market crashes have in common, although the underlying 
physical mechanisms are fundamentally different. 
SOC behavior can also be simulated by mathematical 
rules, as it is demonstrated with cellular automaton models (Pruessner 2012). 
Consequently, there is both (i) an universally valid statistical aspect 
of SOC systems that leads to scale-free power law-like size distributions,
and (ii) physical scaling laws that are specific to each particular SOC 
phenomenon. Physical scaling laws can often be expressed in terms of 
variables that are coupled to each other in a multiplicative way by 
(nonlinear) exponents, from which the power law slopes of the observed
size distributions can be calculated. 

Power law size distributions are the hallmarks of self-organized
criticality (SOC) systems, and therefore the determination of the power law
slope $\alpha$ is an important quest, but power law fitting is not trivial
because of a number of deviations that are not properly understood.
While the standard SOC models call for an ideal power law distribution 
function, it was rigorously proven with maximum likelihood fitting
methods with goodness-of-fit tests based on the Kolmogorov-Smirnov
statistic that most observed size distributions of empirical data
are not consistent with an ideal power law, a power law with a cutoff,
an exponential, a stretched exponential, or a log-normal distribution
function (Clauset et al.~2009). Some sceptics went even as far as to
deny the existence of power laws at all (Stumpf and Porter 2012). 
Here we propose to add a threshold constant to the ideal power law
function, which describes a physical threshold, an instrumental sampling
threshold, or a background contamination, which appears to fit a large 
number of observed (power law-like) size distributions. The introduction 
of a threshold constant allows us to fit an observed size distribution 
over a larger range without having to make any assumption of an arbitrarily 
chosen scale-free range over which a power law can be fitted. We will 
test this concept with terrestrial, solar, and astrophysical data sets,
some of them investigated previously (Clauset et al.~2009;
Aschwanden et al.~2015). We use the term {\sl thresholded power law size
distribution} throughout the paper, which is also known in the statistical
literature as {\sl ``Generalized Pareto distribution''} (Hosking and Wallis 1987), 
``Generalized Pareto Type II distribution'' (e.g., Arnold 2015; Johnson 
et al.~1994), or ``Lomax distribution'' (Lomax 1954).

This paper is organized into a theoretical section on the definition of
thresholded size distributions, for differential and cumulative
occurrence frequency distributions (Section 2), a description of the
data analysis method (Section 3), analysis of terrestrial, solar, 
and astrophysical data sets (Section 4), a discussion of the results 
(Section 5), and conclusions (Section 6).

\section{		THEORY						}

\subsection{		Instabilities in Astrophysical Plasmas		}

Virtually all nonlinear energy dissipation processes in astrophysics 
are governed by some instability, which all have a distinct threshold. 
These systems are stable as long as the conditions remain below some 
critical threshold, while the onset of an instability is triggered when 
a critical threshold is locally exceeded. We list a selection of common 
stabilities that occur in solar and astrophysical plasmas in Table 1, and
visualize their physical mechanisms in Fig.~1, including mostly
hydrodynamic, magneto-hydrodynamic, and kinetic instabilities (for references see
textbooks, e.g., Krall and Trivelpeace 1971; Cowling 1976; Schmidt 1979; Melrose
1980a,b, 1986; Priest 1982; Benz 1993; Sturrock 1994; Kivelson and Russell 1995;
Baumjohann and Treumann 1996; Treumann and Baumjohann 1997; Somov 2000; Priest
and Forbes 2000; Biskamp 2000; Tajima and Shibata 2002; Goossens 2003; 
Aschwanden 2004; Bellan 2006).  

For the onset of an instabiltiy to be triggered (Fig.~2 left), the threshold 
condition could be a critical density gradient (Rayleigh-Taylor instability), 
a critical velocity gradient (Kelvin-Helmholtz instability), a critical 
Lorentz force (ballooning instability, resistive instabilities, current 
pinch instabilities), a critical temperature gradient (thermal instabilities), 
or a critical gradient in the particle velocity distribution (bump-in-tail, 
two-stream, or loss-cone instabilities).

These instabilities play a fundamental role for the generation of most 
astrophysical phenomena, such as the formation of supermassive black holes 
triggered by a global dynamical instability (Begelman et al.~2006), 
the formation of galactic dark matter structures by a gravitational 
instabilty (Springel et al.~2005), star formation by a gravitational
instability (Wang and Silk 1994), giant planet formation by a gravitational 
instability (Boss 1997), solar flares by magnetic reconnection 
instabilities (Priest and Forbes 2000), such as by the kink-mode or 
tearing mode instability, solar radio bursts by wave-particle interactions 
and kinetic instabilities (Benz 1993), coronal loop evolution by thermal 
instabilities (Priest 1978), auroral kilometric radiation bursts from 
planets by the electron-cyclotron loss-cone instability (Wu and Lee 1979),  
or magnetospheric substorms by velocity shear-related instabilities (Lui 1991).

Since instabilities cannot occur without crossing a critical threshold, 
instability thresholds are an essential ingredient in the statistical 
distributions of observed astrophysical (or terrestrial) phenomena. 
In the next section we present a simple analytical model that predicts 
the statistical distribution of phenomena that are subject to an instability 
that operates above some physically defined threshold value. 

\subsection{ 	Thresholded Exponentially Growing Instabilities 	} 

There is a simple analytical model of nonlinear processes in terms
of an exponential growth phase with saturation after a random time
interval, which goes back to Willis and Yule (1922) who applied it to
geographical distributions of plants and animals. Yule's model
was also applied to cosmic rays (Fermi 1949), to cosmic transients
and solar flares (Rosner and Vaiana 1978; Aschwanden et al.~1998),
or to the growth dynamics of the world-wide web (Huberman and
Adamic 1999), as well as to the distribution of the sizes of incomes,
cities, internet files, biological taxa, and gene and
protein families (Reed and Hughes 2002). The time evolution of
an exponentially growing avalanche starts at a threshold value $x_0$
at time $t=0$,
\begin{equation}
	x(t) = x_0 \left[ \exp{\left( {t \over t_G} \right)} - 1 \right] \ ,
\end{equation}
where $t_G$ represents the exponential growth time. The size of the
avalanche is normalized to zero for avalanches starting at time $t=0$.
If the exponential growth process saturates after a random time interval $t$,
the statistical probability can be approximated with an exponential 
distribution function,
\begin{equation}
	N(t) dt = {n \over t_S} \exp{\left( - {t \over t_S} \right)} \ dt \ ,
\end{equation}
where $t_S$ represents the e-folding saturation time. We can now invert
Eq.~(1) to obtain the time dependence, $t(x) = t_G \ln{(x/x_0+1)}$, and
the derivative $dt/dx=(t_G/x_0)(x/x_0+1)^{-1}$, and directly obtain the
size distribution $N(x)$ of exponentially growing avalanches,
\begin{equation}
	N(x) dx = N(t[x]) \left| {dt \over dx} \right| dx
		= {n (\alpha_x - 1) \over x_0} 
		  \left({x \over x_0}+1 \right)^{-\alpha_x} dx \ ,
\end{equation}
where the power law index has the solution 
\begin{equation}
	\alpha_x = \left (1 + {t_G \over t_S} \right) \ .
\end{equation}
A more detailed discussion of this exponential-growth model can be found in 
Aschwanden (2011c; \S 3.1). 

The size distribution $N(x)$ specified in Eq.(3) can be written as,
\begin{equation}
	N(x) dx = n_0 \left( x_0 + x \right)^{-\alpha_x} dx \ ,
\end{equation}
with the normalization constant $n_0$ for the range $x_1 \le x \le x_2$,
\begin{equation}
	n_0	=n_{ev} (1-\alpha_x) 
		\left[ (x_2+x_0)^{1-\alpha_x}-(x_1+x_0)^{1-\alpha_x}
		\right]^{-1} \ .
\end{equation}
This size distribution (Eq.~5) is identical to a 
{\sl ``Generalized Pareto distribution''} (Hosking and Wallis 1987), 
the ``Generalized Pareto Type II distribution'' (e.g., Arnold 2015; 
Johnson et al.~1994), or ``Lomax distribution'' (Lomax 1954),
to which we refer as {\sl thresholded power law size distribution} in the
remainder of this Paper.
Such a distribution function is also called a {\sl differential
occurrence frequency distribution function}, a {\sl log(N)-log(S)}
diagram, or simply a {\sl size distribution}. 

The ideal power law (or Pareto) distribution function of a parameter $x$,
with a power law coefficient $\alpha_x$ that extends over a
scale-free range of $x_1 \le x \le x_2$, can be retrieved from the
generalized Pareto distribution simply by setting $x_0=0$ in Eq.~(5)
\begin{equation}
        N(x) dx = n_0 \ x^{-\alpha_x} dx \ , \qquad x_1 \le x \le x_2 \ ,
\end{equation}
with $n_0$ being the normalization constant,
\begin{equation}
        n_0     =n_{ev} (1-\alpha_x) \left[ x_2^{1-\alpha_x}-x_1^{1-\alpha_x}
                \right]^{-1} \ .
\end{equation}

\subsection{ 	Cumulative Distribution Functions			}

For small samples, a {\sl cumulative distribution function (CDF)} is often fitted 
instead, because the small number of events is not sufficient to bin the data.
A cumulative size distribution is simply defined by the integral
of the total number of events above a given value $x$, where
$x_1$ represents the minimum value and $x_2$ the maximum value
of the size distribution, over which the CDF is integrated.
For the ideal power law distribution function (Eq.~7) we have,
\begin{equation}
	N_{cum}(>x) dx = \int_{x}^{x_2} n_0\ x^{-\alpha_x} dx
	= 1 + (n_{ev}-1) \left( { x_2^{1-\alpha}-x^{1-\alpha} \over 
	                  x_2^{1-\alpha}-{x_1}^{1-\alpha} } \right) \ .
\end{equation}
where $n_{ev}$ is the total number of events included in the size distribution.
Strictly speaking, this definition is only valid for exponents 
$\alpha_x \neq 1$, while the integral takes on the form of $\propto \ln(x)$
for $\alpha_x=1$.
Note that this distribution has an asymptotic value of $N_{cum}(>x_1)=n_{ev}$
at the lower bound, and vanishes at the upper bound, $N_{cum}(>x_2)=0$. 
It is important to note that the cumulative size distribution (Eq.~9)
is not exactly a power law function with
a slope of $\beta = \alpha-1$, but rather shows a steep drop-off at the
upper end like an exponential function, due to the highest value that
forces the probability to vanish above $x_2$.
It is a consequence of the fact that the differential size distribution
has a sharp upper bound at $x_2$ and does not continue to infinity, an
effect that is also called a {\sl finite system size effect} in
SOC models. This deviation from a straight power law (of the 
cumulative size distribution) is often ignored, or is modeled with an 
{\sl ad hoc} (exponential) term, i.e., $N(x) dx \propto x^{-\alpha_x}
exp(-x/x_0)$ (e.g., Lu et al.~1993), but we will fit it
properly to the data in this study by fitting the exact analytical
distribution function defined in Eq.~(9) (see examples in Figs.~3c and 3d).

The thresholded size distribution has the follwoing CDF,
\begin{equation}
	N_{cum}(>x) = \int_{x}^{x_2} n_0 
	( x + x_0 )^{-\alpha_x} dx 
	= 1 + (n_{ev}-1) \left( {(x_2+x_0)^{1-\alpha_x}-(x  +x_0)^{1-\alpha_x} \over 
	                 (x_2+x_0)^{1-\alpha_x}-(x_1+x_0)^{1-\alpha_x} } \right) \ .
\end{equation}
The bounds $[x_1, x_2]$ and the number of events $n_{ev}$ are known in principle,
which leaves two free variables ($x_0, \alpha$). From the definition given in 
Eq.~(10) we see immediately that the cumulative size distribution obtains the 
value $N_{cum}(>x_1) = n_{ev}$ at the lower bound $x=x_1$, and 
$N_{cum}(\ge x_2) = 1$ at the upper bound $x=x_2$. A set of differential
size distributions $N(x)$ (for $x_0=0, 10, 20, ..., 50$) is shown in Fig.~(3b),
as well as the corresponding set of cumulative
size distributions $N(>x)$ (Fig.~3d). Further examples of such
size distributions with different power law slopes $a_x=1.0, 1.2, ..., 3.0$
are shown in Fig.~(4b) and (4d).

\subsection{ 	Truncation Effects due to Incomplete Sampling 	}

The low end of a power law-like distribution function often shows 
a deviation from an ideal power law function in form of a
gradual rollover, which is caused by undersampling below the 
detection threshold. The detection threshold is a different 
physical cause than the instability threshold $x_0$ derived in
Section 2.2, but can be mathematically treated in the same way.
Therefore, the (positively-defined) parameter $x_0 > 0$ 
in the thresholded size distribution
(Eq.~5) has a double meaning, it could be caused by the physical 
threshold of the relevant instability, or it could be caused by 
the finite sensitivity of an instrument. This ambiguity can only
be resolved by using additional information. For instance, if a
data set with higher sensitivity is used, the value $x_0$ should
move to lower values (Fig.~2 right), 
while it should stay constant for a physical
instability threshold that is instrument-independent 
(see effect of instability threshold $E_{T0}$ and detection 
threshold $E_{D0}$ on cumulative size distribution in Fig.~2 right panel). 

\subsection{ 	Background Contamination }

Actually, the parameter $x_0$ has a third meaning, if it is negative,
$x_0 < 0$. 
In any size distribution of a certain type of events, it is always
possible to have an event-unrelated background that is responsible for false
detections at the level of the smallest events. In particular 
in astrophysical data sets, the flux $f(t)$ of events that is sampled 
may often contain some event-unrelated background $b(t)$, so that the
total observed flux $F(t)=f(t)+b(t)$ represents an over-estimate of
the event-related flux $f(t)$. For instance, hard X-ray pulses originating
from solar flares always contain some background photons of the same energy
that originate from non-solar sources (such as the galactic background).
Such a background contamination effect has the same functional form as 
a thresholded power law distribution function (Eq.~5),
\begin{equation}
	N(x) dx = n_0 ( x + x_0 )^{-\alpha_x} \ dx 
	        = n_0 ( x - b_0 )^{-\alpha_x} \ dx \ ,
\end{equation}
except that the offset constant $x_0 = - b_0$ is negative. 
A set of differential size distributions $N(x)$ (for $x_0=0, 10, 20, 
..., 50$) is shown in Fig.~(3a), as well as the 
corresponding set of cumulative size distributions $N(>x)$ (Fig.~3c). 
Further examples of such size distributions for 
different power law slopes $a_x=1.0, 1.2, ..., 3.0$ are shown in Fig.~(4a)
and (4c).  We see that the lower end of the size distributions
show a steepening of the power law slope, while a cutoff occurs at 
$x=b_0$. In practice, the background $b_0$ may not be a constant value
for each event, but for sake of simplicity we study here only the 
zero-order effect of a constant background value.

\section{	NUMERICAL METHODS 				}

After we described the theoretical framework of 
thresholded power law distributions (Section 2.2, 2.3), and the
effect of incomplete sampling (Section 2.4) and background contamination 
(Section 2.5), we turn now to the
numerical method to fit these modified power law distribution functions
to data sets and to evaluate a goodness-of-fit criterion. We will use
Monte-Carlo simulations to validate the numerical method used here.

\subsection{ Monte-Carlo Simulations of Power Law Distributions }

We start with Monte-Carlo simulations of generalized (thresholded)
power law distributions of the form of Eq.~(5). We wish to generate
a data set $x_i, i=1,...,n$ as a function of random numbers $\rho_i$
that are drawn from a random generator in the range of $\rho_i=[0,1]$,
and produce a size distribution function $N(x)$ as prescribed by Eq.~(5).
The general procedure to set up such a Monte Carlo simulation is 
described in some detail in Clauset et al.~(2009; Appendix D) or 
Aschwanden (2011c; \S 7.1.4). First we normalize the differential probability 
function to unity,
\begin{equation}
	p(x) = {N(x) \over n_{ev}} = {n_0 \over n_{ev}}
	( x + x_0 )^{-\alpha_x} \ ,
\end{equation}
and obtain the total probability distribution function $P(x)$ integrated 
over the range $[x_1,x]$,
\begin{equation}
	P(x) = \int_{x_1}^x \ p(x')\ dx' 
	     = \left( {(x  +x_0)^{1-\alpha_x}-(x_1+x_0)^{1-\alpha_x} \over 
	               (x_2+x_0)^{1-\alpha_x}-(x_1+x_0)^{1-\alpha_x} } \right) ,
\end{equation}
and set it equal to the (normalized) sorted random values $\rho_i=[0,1]$,
\begin{equation}
	P(x) = \int_{x_1}^x \ p(x')\ dx' = \int_0^\rho d\rho' = \rho \ ,
\end{equation}
which yields a relationship between the sizes $x_i$ and random values $\rho_i$
and can explicitly be expressed as a function $x(\rho)$ by the inversion of 
the total probability distribution function (Eqs.~13, 14),
\begin{equation}
	x(\rho) = \left[ \rho (x_2+x_0)^{1-\alpha_x} 
	+ (1 - \rho) (x_1+x_0)^{1-\alpha_x} 
	\right]^{1/(1-\alpha_x)} - x_0 \ .
\end{equation}
This numerical procedure allows us to simulate the generalized
power law distribution functions with different random number sets,
by using a random generator to produce values $\rho_i=[0,1]$ that map
into the size range $x_i=[x_1,x_2]$ and reproduce the probability
distribution function of Eq.~12). Such Monte-Carlo simulations
provide uncertainties of the power law size distribution
parameters ($\alpha_x, n_0, x_0$) due to random noise, which can be
used to calculate a $\chi^2$ goodness-of-fit criterion that tells us 
which of the observed distributions are consistent with the theoretical 
model of a thresholded power law distribution function (Eq.~5).

\subsection{ 	Least-Square Goodness-of-Fit Test }

We simulate 10 different data sets with $x_0=10$, $a=2.0$, with different
numbers of events, $n_{ev}=10^5 \times 2^{(-i)}$, $i=1,...,10$. Each one
has uniformly distributed random numbers $\rho_i$, $i=1,...,n_{ev}$ from 
a range of unity $\rho_i=[0,1]$. We transform the values $\rho_i$ into
size variables $x_i(\rho_i)$ according to the function of Eq.~(15),
and sample them in $n_x=40$ uniform logarithmic bins in the range of
$x_i=[1,10^5]$. We show the resulting differential size distributions 
$N(x)$ (Fig.~5a) and cumulative size distributions (Fig.~5c).
The counted number of events per bin in the differential size
distribution function is $N_{bin,i}$, while the differential distribution 
function is defined by dividing with the (non-equidistant) bin width 
$\Delta x_i$, so that $N_i(x_i)=N_{bin,i}(x_i) / \Delta x_i$. The
expected uncertainty of the differential size distribution is then
\begin{equation}
	\sigma_{diff,i}=\sqrt{(N_i \Delta x_i)} / \Delta x_i \ ,
\end{equation}
while the expected uncertainty of the cumulative size distribution,
which we represent with the same (logarithmic) binning, is
\begin{equation}
	\sigma_{cum,i}=\sqrt{N_i} \ .
\end{equation}
We can now calculate a goodness-of-fit $\chi^2$-criterion for the
difference between the simulated distributions $N_{sim,diff}(x)$ and the
theoretical distribution function $N_{fit,diff}$ for both the differential
size distribution function (with $n_{par}=3$ parameters, $a, x_0, n_0$),
\begin{equation}
	\chi_{diff} = \sqrt{ {1 \over (n_x-n_{par})} 
	\sum_{i=1}^{n_x} { [N_{fit,diff}(x_i) - N_{sim,diff}(x_i)]^2
	\over \sigma_{diff,i}^2 } } \ ,
\end{equation}
and the cumulative distribution function $N_{fit,cum}(>x)$,
\begin{equation}
	\chi_{cum} = \sqrt{ {1 \over (n_x-n_{par})} 
	\sum_{i=1}^{n_x} { [N_{fit,cum}(x_i) - N_{sim,cum}(x_i)]^2
	\over \sigma_{cum,i}^2 } }\ ,
\end{equation}
which are both shown in Figs.~(5b) and (5d), together with error
bars based on 1000 different realizations of random number sets. This
test shows clearly that the defined goodness-of-fit criterion yields 
the expected value of $\chi^2 \approx 1.0$ for pure random noise.
The statistical averages are $\chi^2=1.05 \pm 0.02$ for the
differential size distributions (Fig.~5b), and $\chi^2=0.82\pm0.01$
for the cumulative size distributions (Fig.~5d). The latter value deviates 
from the expectation value of $\chi^2$ due to the statistical dependence of
the counts in binned cumulative size distribution, but is close enough
to unity to serve as a sensible goodness-of-fit criterion.

Note that the events in each bin of the differential size distribution 
are statistically independent, while the number of events $N_{cum}(>x)$ in
the cumulative size distribution are statistically dependent. However,
since we use a fixed (logarithmic) binning, the fluctuations of events
in each bin, $N_{cum}(>x)=\int_x^{x_2}$, obey approximately
the same random statistics
$\sigma_{cum,i}=\sqrt{N_{cum,i}}$ in each bin as for the differential size
distribution, $\sigma_{diff,i}=\sqrt{N_{diff,i}}$, so that we have 
uncertainty estimates $\sigma_{diff,i}$ (Eq.~16) and $\sigma_{cum,i}$ 
(Eq.~17) for both distributions and can apply the goodness-of-fit 
critera $\chi_{diff}$ (Eq.~18) and $\chi_{num}$ (Eq.~19) for both types 
of distributions, as proven in Fig.~(5b) and (5d).

\subsection{ 	Fitting of Power law Distributions }

In the following data analysis, the simulated distributions will be
replaced by the distributions of the observed data, 
$N_{sim}(x) \mapsto N_{obs}(x)$,
while the theoretical distribution function will be replaced by the actual
best fit of the theoretical model, $N_{fit}(x)$. Calculating the
goodness-of-fit $\chi^2$ criteria (Eqs.~18, 19) will tell us whether 
the data are consistent with the theoretical model (Eq.~5) or not. 

Alternative methods to the least-square fit test are maximum likelihood
estimators for generalized Pareto distributions (Goldstein 2004; 
Newman 2005; Bauke 2007; Giles et al.~2011), the Bayesian 
information criterion (BIC), the Kolmogorov-Smirnov (KS) test, or the 
Anderson-Darling (AD) test, which have been applied to the cumulative 
size distributions of empirical data sets in Clauset et al.~(2009). 
In the latter application, 
the KS test was considered to yield good results, while the AD test was 
found to give estimates of the lower bound $x_1$ (of the power law
range in the cumulative size distribution) that are too large by an 
order of magnitude or more, while the BIC method was found to have 
a tendency of underestimating $x_1$ (Clauset et al.~2009). Since
our Monte-Carlo simulations validated the $\chi^2$ minimization
goodness-of-fit test for both the differential (Fig.~5b) and 
cumulative size distribution (Fig.~5d), we will use this method 
throughout this study.	  

\medskip
Our data analysis procedure of fitting thresholded power law size
distributions to the data and evaluating the goodness-of-fit comprises
the following steps: (1) histogram binning of both the differential 
and cumulative size distributions; (2) determination of the threshold 
value $x_0$; (3) optimization of the free parameters $n_0$ and $a$
and least-square goodness-of-fit test; and (4) elimination of 
event-unrelated background contamination. We briefly describe these
four steps in turn.

\underbar{Step 1: Histogram binning:} The input from every data set 
is a series of event sizes $x_i, i=0,...,n_{ev}$
with a total of $n_{ev}$ events. This series is defined by a minimum 
$x_1=min(x_i)$ and maximum value $x_2=max(x_i)$.
We apply a logarithmic binning to the data for both the differential
and the cumulative size distributions, which is uniform on a logarithmic 
scale and covers the sampled data range $[x_1, x_2]$.
For differential size distributions we use $n_{dec}=10$ logarithmic
bins per decade, yielding $n_{bin}=^{10}log(x_2/x_1) \times n_{dec}$ bins 
per data set, while we use $n_{dec}=5$ logarithmic bins per decade for
cumulative size distributions. Bins without events in the differential
size distribution (which show no change in the number of events in the
cumulative size distribution) are ignored.

\underbar{Step 2: Threshold $x_0$:} Since the lower end of the
(differential or cumulative) size distribution is often subject to
incomplete sampling, it is necessary to find a lower bound or
threshold $x_0$ that defines an upper range $[x_0, x_2]$ where
events are completely sampled and a power law fit can be applied.
We find that a reliable lower bound $x_0$ can be determined from
the bin where the number of events per bin has a maximum,
\begin{equation}
	N_{diff,max} = max[N_{diff}(x)] = N_{diff}(x=x_0) \ .
\end{equation}
We show an example in Fig.~(6), where the peak counts of a data set
of 10,665 solar flares are sampled in form of a logarithmically binned 
histogram (with data taken from Clauset et al.~2009). 
The number of events $N_{bin,i}$ per (logarithmic) bin 
is shown in Fig.~(6a), where the maximum of the distribution 
defines the threshold value $x_0$. The resulting differential size
distribution is obtained from dividing the number of events per bin
by the (logarithmic) bin width, $N_{diff}(x)=N_{bin,i}/\Delta x_i$
(Fig.~6b), and the cumulative distribution function is
obtained from the the number of events in the integrated ranges
$[x,x_2]$, namely $N_{cum}(>x_i)=\int_x^{x_2} N_{diff,i}(x) dx$
(Fig.~6c). 

The rationale for this method of finding a lower threshold $x_0$
is justified as follows. For an ideal power law distribution that 
is completely sampled over the entire range $[x_1,x_2]$, we would 
expect that the number of events per bin monotonically increases 
towards the lower bound, $x \mapsto x_1$, for power law slopes steeper
than $a > 1$, such as shown in the simulation for $a=2.0$ in Fig.~(5c.) 
In real data, the lower range is generally 
sampled incompletely (unless we have a 100\% detection efficiency
all the way down the size of the smallest event in the data set). 
Therefore, if we find a decrease in the number of events per bin when
moving towards the lower bound, say in the range $x_1 \le x \le x_0$, 
this decrease in events indicates undersampling, or an incompletely 
sampled range, and thus it is appropriate to limit the fitting range 
to the completely sampled range $[x_0,x_2]$.

\underbar{Step 3: Optimization of the free parameters $n_0$ and $\alpha$:}
Fitting the theoretical function (Eq.~5) to the data involves three 
free variables, the threshold $x_0$, the normalization constant $n_0$, 
and the power law slope $a$. In principle, the constant $n_0$ is known 
from the normalization to the total number of events in the data set,
$n_{ev}$ (Eq.~6), but this is only true for completely sampled data 
sets where the threshold coincides with the data minimum value
($x_0=x_1$), such as in the data simulations shown in Figs.~3 and 4. 
Since the threshold $x_0$ is already
determined in the previous step, we have only two free variables ($n_0$ 
and $\alpha$) to optimize, which we obtain by means of minimizing the 
$\chi^2$-values (Eqs.~18 and 19) that express the least square
difference between the theoretial model $N_{fit}(x_i)$ (Eq.~5) 
and the histogrammed data points $N_{obs}(x_i)$, $i=0,...,n_{bin}$.
We perform the optimization of the $\chi^2$-values by using a Powell
minization method (Press et al.~1986). The resulting $\chi^2$-values 
provide also a goodness-of-fit test, whether the model fits the data
(in case of $\chi^2 \approx 1$). Values of $\chi^2$ significantly
above unity indicate a failure of the model fit, while values 
significantly below unity indicate noisy data that are not sensitive 
to the model fit.

Note that the fitted parameter $a$ corresponds to an effective slope 
$b=a-1$ for the cumulative size distribution. We estimate 
an error $\sigma_a$ of the fitted power law slope $a$ from a relationship 
that scales with the square root of the number $n$ of events and was
derived earlier from Monte-Carlo simulations (Aschwanden 2011a),
\begin{equation}
	\sigma_\alpha = {\alpha \over \sqrt{n} } \ .
\end{equation}
which is slightly different from the definition obtained with maximum
likelihood estimators in generalized Pareto distributions (e.g., in
Clauset et al.~2009), $\sigma_{\alpha}=(\alpha-1)/\sqrt{n}$.
Thus we expect an accuracy of approximately 10\% for $n=100$ events, or
1\% for $n=10^4$ events. 

\underbar{Step 4: Elimination of event-unrelated background:}
If the data reveal a steepening of the power law slope near the threshold
value $x_0$, such as shown in Fig.~(3a) or (3c),
this indicates the presence of event-unrelated
background contamination. To first order, we can correct for this
effect by subtracting a suitable background value $b_0$, i.e.,
$x_{corr}=x - b_0$. If the background value $b_0$ would be an exact
constant, a sharp lower cutoff at $x=b_0$ would occur in the size
distributions, such as shown in the Monte-Carlo simulations in Figs.~(3a,3c)
or in Figs.~(4a,4c). In reality, there are
events with values of $x < b_0$ in the sample, which become negative
after correction and get lost from a positively defined size distribution.
Since we are only interested in a corrected power law slope above
a threshold value $x_0 > x_b$, we ignore negative values after correction,
but we have to be aware that this reduces the overall number of 
sampled events.  

In the absence of information on the correct background value for each
event, the question arises how can we determine a mean value 
$b_0$ empirically? As the Monte-Carlo simulations in Figs.~3 and 4 show, 
the size distributions with a steepening at the lower end converge to 
the extrapolated power law slope, when the correct background is subtracted. 
Therefore, we can find the correct mean background value by varying the values
$x_0=-b_0$ in the fitted thresholded power law function (Eq.~5) until 
the $\chi^2$-values reach a minmum. We demonstrate this procedure
in Fig.~(7). A differential size distribution is shown for an observed
data set of 10,665 solar flares (taken from Clauset et al.~2009), 
with no background subtraction (Fig.~7a), 
which shows a clear discrepancy between the best fit
of a thresholded power law function and the power law slope at the
upper bound, which is interpreted as an effect of unsubtracted background 
that steepens the distribution at the lower end. The resulting goodness-of-fit
with no background subtraction, $\chi^2 \approx 6$ (Fig.~7d), 
clearly indicates a bad fit. If we increase the level of background
subtraction, the goodness-of-fit improves and reaches a best value of
$\chi^2 \approx 1.0$ at a background level of $b_0=57$ cts s$^{-1}$
(Fig.~7b and 7d).
The goodness-of-fit remains at a similar level for subtraction of larger
backgrounds ($b_0=57-120$), because no background contamination occurs
at larger events than $x \ge b_0$. After we have determined the correct
background $b_0$, step 3 is repeated to obtain optimum values for
the parameters $n_0$ and $a$ of the background-subtracted values 
$x_{corr} = x - b_0$. 

\section{	DATA  ANALYSIS 			}

We are fitting our analytical model of differential and cumulative
thresholded size distributions to a number of data sets from terrestrial,
solar, and stellar data. The mostly terrestrial data stem from Clauset 
et al.~(2009), the solar data have been previously reviewed in 
Aschwanden et al.~(2015), and the stellar data are obtained from the 
{\sl Kepler} mission, either from published or newly processed data.  
We analyze all size distributions with the same unified methodology
and compare the obtained power law slopes with published values 
obtained with alternative methods.

\subsection{	Terrestrial and Empirical Data				}

We use 9 data sets out of the 24 empirical data sets published
by Clauset et al.~(2009) for two reasons: (1) Most of these data sets 
have large statistics in the order of $10^4-10^5$ events, which is
needed for accurate power law fits, and (2) nine of these
data sets are accessible from public websites (see URL links,
description of data sets, and origin with references in Clauset
et al.~2009). For all of these data sets, ideal power law
functions (Eq.~7) have been fitted, a lower bound $x_0$ for
the power law range has been determined  and the power law hypothesis
has been tested in Clauset et al.~(2009). Alternatively, we fit here
{\sl thresholded power law distribution functions} (Eq.~5) to both
differential (Fig.~8) and cumulative size distributions (Fig.~9).
In addition we test the hypothesis whether thresholded power laws
are consistent with the data with the $\chi^2$ goodness-of-fit 
critera. The results are summarized in Table 2.

The 9 empirical data sets selected from Clauset et al.~(2009) are
gathered from one solar (a), two geophysical (b,c), three social (d,e,f), 
and three information-type (g,h,i) statistical data sets:
(a) hard X-ray peak count rates from {\bf solar flares} observed with
HXRBS/SMM (Dennis 1985; Schwartz et al.~1992; Crosby et al.~1993;
Newman 2005); 
(b) the intensities of {\bf earthquakes} occurring in
California between 1910-1992 measured from the maximum amplitude
of motion (Newman 2005); 
(c) the areas of {\bf wildfires} (in units of acres)
occurring on USA federal land between 1986 and 1996 (Newman 2005);
(d) the human population of USA {\bf cities} in the 2000 US Census; 
(e) the number of customers affected in electrical {\bf blackouts} 
in the USA during 1984-2002 (D'Agostino et al. 1986); 
(f) the casualties of {\bf terrorist} attacks worldwide from
February 1968 to June 2006 (Clauset et al.~2007); 
(g) the frequency of occurrence of unique
{\bf words} in the novel {\sl Moby Dick} by Herman Melville (Newman 2005); 
(h) the frequency of occurrence of US
family {\bf surnames} in the 1990 US Census; and 
(i) the number of {\bf web links} to web sites found in a 1997 web crawl 
of about 200 million web pages (Broder et al.~2000). 

If we allow a generous limit of $\chi^2 \lapprox 1.3$ for the 
goodness-of-fit in the cumulative size distribution, we find that 
4 out of the 9 empirical data sets are consistent with the hypothesis
of a thresholded power law size distribution (flares, quakes, blackouts,
surnames), while the other 5 cases are inconsistent with the hypothesis
(fires, cities, terrorism, words, weblinks). Clauset et al.~(2009) finds support 
for a power law, or for a power law with cut-off, for all of these 9 cases, 
but the specific shape of the cutoff is not fitted and the lower bound
$x_0$ is determined differently than in our work.

When we look at the power law slopes and their uncertainties (Table 2), 
we find good agreement between our differential and cumulative size 
distributions within the quoted uncertainties in almost all cases
(Table 2: 5th and 6th column).
Comparing with the power law slopes obtained by Clauset et al.~(2009),
we find approximate agreement (within the stated uncertainties) 
for a few data sets only (flares, words, surnames, weblinks).
The remaining discrepancies are mostly caused by the choice of the 
fitting range, which is dictated by the threshold value $x_0$ of
complete sampling in our study, while Clauset et al.~(2009) use
a heuristic method.
Those cases that are not consistent with the thresholded 
power law model, such as the cities and fires, exhibit significant 
deviations over a large portion of the fitting range, where a power law 
fit is ill-defined anyway.

Regarding background subtraction we found the largest contamination
for the data set of solar flares ($b_0=57$ cts s$^{-1}$), which 
includes the preflare background count rate. This is
actually fairly consistent with the galactic hard X-ray background
count rate for the HXRBS/SMM data set, which is estimated to produce
a mean background level of $\approx 40$ cts s$^{-1}$ (Dennis 1985).
A minor background component of 1-2 units was found also for five
other data sets (fires, terrorism, quakes, words, surnames, weblinks), 
which could possibly be attributed to a discretization effect (since 
the lowest values in these data sets are digitized by discrete integer 
numbers).

In summary, we find a good consistency between the power law slopes 
determined from the differential and cumulative size distributions,
but we find substantial differences between the power law slopes
determined with our $\chi^2$-square method and the {\sl maximum-likelihood
fitting (MLF)} method of Clauset et al.~(2009), which {\bf is caused} 
by differently chosen fitting ranges, rather than by the fitting
method. The choice of the distribution model (e.g., ideal power law
versus the thresholded power law model) appears to be more important
in determining which model is most consistent with the data than the
fitting method.

\subsection{	Solar Data					}

One of the most suitable data sets with power law behavior in solar
physics are count rates of hard X-ray peak fluxes $P$ from solar flares,
because of their near-perfect match to a power law function, their 
large statistics ($\approx 10^3-10^5$ events per data set), multiple
instrument coverage (SMM, CGRO, RHESSI), multiple solar cycle coverage
(Cycle 21-24), and multiple published analyses.

Solar flares provide the energy source for acceleration of nonthermal
particles, which emit bremsstrahlung in hard X-ray wavelengths, once
the non-thermal particles interact with a high-density plamsa via
Coulomb collisions. Most solar flares display an impulsive component
in hard X-rays, produced by accelerated coronal electrons that
precipitate towards the chromosphere and produce intense hard X-ray
emission at the footpoints of flare loops. Therefore, hard X-ray
pulses are a reliable signature of solar flares, often detected
at energies $\gapprox 20$ keV, but for smaller flares down to
$\gapprox 8$ keV.

Solar flare event catalogs containing the peak rate $(P)$,
fluences $(E)$, and flare durations $(T)$, have therefore been
compiled from a number of spacecraft or balloon-borne hard X-ray
detectors over the last three decades, such as from OSO-7
(Datlowe et al.~1974), a University of Berkeley balloon flight
(Lin et al.~1984), HXRBS/SMM (Dennis 1985; Schwartz et al.~1992; Crosby
et al.~1993), BATSE/CGRO (Schwartz et al.~1992; Biesecker et
al.~1993, 1994; Biesecker 1994), WATCH/GRANAT (Crosby 1996;
Georgoulis et al.~2001); ISEE-3 (Lu et al.~1993; Lee et al.~1993;
Bromund et al.~1995); PHEBUS/GRANAT (Perez-Enriquez and
Miroshnichenko 1999).  RHESSI (Su et al.~2006; Christe et al.~2008;
Lin et al.~2001), and ULYSSES (Tranquille et al.~2009).
A compilation of occurrence frequency distribution power law slopes
of solar hard X-ray flare peak fluxes ($\alpha_P$), fluences
or energies ($\alpha_E$), and flare durations ($\alpha_T$) is listed
in Table 3. 

We show the fits of differential and cumulative size distributions
to the HXRBS/SMM data set in Figs.~10 and 11 (top panels). The entire 
data set comprises 10291 events from the 1980-1989 (solar cycle 21/22) 
period, but the subtraction of an average background of 57 cts s$^{-1}$ 
reduces the sample of peak fluxes to 7931 events. The results for the 
BATSE/CGRO data set from 7233 flares observed during the epoch of 
1991-2000 (solar cycle 22/23) are shown in Figs.~10 and 11 (middle panels), 
which reveals no significant background (in the catalogued values). 
The results for the RHESSI 
data set from 11549 flare events observed during 2002-2010 (cycle 23/24) 
is shown in Figs.~10 and 11 (bottom panels), where we found an empirical 
background of 35 cts s$^{-1}$. For each of the three data sets we
sampled three different parameters, the peak flux $P$, the total
(time-integrated) counts $E$, and the time duration $D$, each one
shown in Figs.~10 and 11. A graphic visualization of the power law slopes
of the 3 solar flare parameters as a function of the size of the
data set is shown in Fig.~12. We notice that most of the power law
slopes are centered in the ranges of $\alpha_P \approx 1.6-1.9$,
$\alpha_E \approx 1.4-1.7$, and $\alpha_D \approx 2.0-3.0$.
The peak fluxes and total counts show generally a more pronounced
power law function than the durations. 

Since there is a spread of values for the power law slopes, even for
the same data set analyzed by different authors (see Table 3), we
investigate a few systematic effects. The variation of fitted power law 
slopes seems not to depend on the size of the data set, if at least
$n \gapprox 10^2$ events are sampled, as it can be seen in Fig.~12.
We varied the level of background subtraction from 0 to 100 cts/s
and find a strong variation of the power law slope for differential
size distributions, while the cumulative size distributions appears
to be less dependent on the background subtraction level
(Fig.~13, top). If we vary 
the number of sampled flare events, we see good agreement between the 
slopes of cumulative versus differential size distributions for data sets
that have a minimum size of $n \gapprox 10^3$ (Fig.~13, bottom),
according to the expected accuracy of the power law slope that scales
as $\sigma_\alpha = \alpha / \sqrt{n}$ (Aschwanden 2011a).
A slight variation of the power law slope during three solar cycles 
was noted also (Fig.~9 in Aschwanden 2011a), which can be explained 
with a variation of the energy threshold for flaring (Aschwanden 2011b).

\subsection{	Stellar Data					}

The capability of obtaining statistics of stellar flare data has been
drastically enhanced recently with observations from the {\sl Kepler
Mission} (Koch et al.~2010), which provides high-precision lightcurves
of individual stars in optical wavelengths. These lightcurves contain
quiescent emission from a star that causes a slow sinusoidal modulation
due to the star rotation, while fast bursts from stellar flares are
superimposed that can accurately be measured after subtraction of the
slowly modulated background (Walkowicz et al.~2011). Searches of
white-light flare emission yielded data of 373 flaring cool dwarf 
stars in the Kepler Quarter 1 data (Walkowicz et al.~2011), 
365 superflares from slowly rotating solar-type stars (Maehara et al.~2012),
1547 superflares on 279 G-type dwarfs (Shibayama et al.~2013), 
209 flares from K-M flare stars and A-F stars (Balona 2015),
and 4944 superflares on 77 G-type stars (Wu, Ip, and Huang 2015).
We tabulate a list of power law slopes that have been measured from
the size distributions of white-light flares observed with Kepler in
Table 4, which are found in the range of $\alpha \approx 1.5-3.0$. 

Our primary interest here is to test the hypothesis whether stellar flare
data are consistent with our hypothesis of thresholded power law size
distributions (Eq.~5). We show the differential size distributions of
three large Kepler data sets ($n_{ev} \gapprox 10^3$) and six smaller
data sets from individual stars in Fig.~14, and the corresponding
cumulative size distributions in Fig.~15. We find that the cumulative 
size distributions are all consistent with the thresholded power law 
size distribution model, based on a goodness-of-fit of $\chi^2 \lapprox 1.0$.
The size distributions of flares from individual stars contain less statistics.
We extracted the six largest data sets, which have 37$-$57 flare events per
star. Those data sets have larger uncertainties and provide less constraints
to test the accurate power law shape. Unfortunately we were not able to
find a star for which we could compare the power law slopes given in
publications and calculated in this study. Nevertheless, the
compilation of analyzed values for the power law slopes of stellar
flares observed with Kepler (Table 4) shows similar ranges among
different studies: $\alpha \approx 2.0-2.3$ in Maehara et al.~(2012),
$\alpha \approx 2.0-2.2$ in Shibayama et al.~(2013),
$\alpha \approx 1.6-2.1$ in Wu et al.~(2015), and
$\alpha \approx 1.5-3.2$ in this study. 
These values for the power law slopes obtained for stellar flares
are systematically higher than for the solar flares tabulated in Table 3,
but we have to be aware that the white-light (bolometric) fluxes are
measured in stellar flares with Kepler, while hard X-ray fluxes
are measured in solar flares with HXRBS, BATSE, and RHESSI,
which may be related to each other by a nonlinear scaling relationship.
Application of physical models to these measurements are beyond the
scope of this paper and will be conducted elsewhere.

Considering the relatively small number statistics of stellar flares,
the measurement of power law slopes is challenging, since the rollover
due to undersampling causes a flattening at the lower end of the 
distribution, and the finite-size effect causes a steepening at the
upper end of the distribution, and there is almost no intervening
range where the power law slope can be reliably measured. It is therefore
imperative to model and fit all three effects simultaneously.

\section{	DISCUSSION				}

\subsection{	The Problematics of Power Laws		}

Stumpf and Porter (2012) write in their article {\sl Critical Truths
About Power Laws}: {\sl``A striking feature that has attracted
considerable attention is the apparent ubiquity of power law relationships
in empirical data. However, although power laws have been reported in areas
ranging from finance and molecular biology to geophysics and the Internet,
the data are typically insufficient and the mechanistic insights are almost
too limited for the identification of power law behavior to be scientifically
useful.}'' This sceptical remark echoed similar thoughts made in an extensive
study on {\sl Power-Law Distributions in Empirical Data} (Clauset et al.~2009),
from which we quote: {\sl Unfortunately, the detection and characterization
of power laws is complicated by the large fluctuations that occur
in the tail of the distribution -- the part of the distribution representing
large but rare events -- and by the difficulty of identifying the range over
which power law behavior holds. Commonly used methods for analyzing power law
data, such as least-square fitting, can produce substantially inaccurate
estimates of parameters for power law distributions, and even in cases where
such methods return accurate answers they are still unsatisfactory because
they give no indication of whether the data obey a power law at all.}''
Clauset et al.~(2009) combine maximum-likelihood fitting methods with
goodness-of-fit tests based on the Kolmogorov-Smirnov statistic and
likelihood ratios, and find that the hypothesis of a power law is consistent
with data in some cases, while it is ruled out in other cases. 

Given these problematics, we identify three major problems: (1) Insufficient
statistics, (2) the definition of the fitting range, and (3) the choice
of the model. The first problem
can be overcome by using more sensitive instruments that can sample events
down to a lower level, or by increasing the observing time span, which 
boosts the number of detected events proportionally to the total duration
of observations. Both trends are ameliorated currently,
because high-technology developments provide us with more sensitive 
instruments in almost all areas, and data sets are now available over
decades of years, since automated measurements started around 1950. 
In particular for solar and astrophysical measurements, the beginning of
the space age after 1956 has opened up the capabilities to sample large
data sets with space-borne instruments in virtually all wavelengths.

The second problem of defining a fitting range for power law distributions
represents a methodical problem that can be solved systematically, as we 
demonstrate in this paper. It is true that most power law fits in the past
have been carried out by eye-balling a straight portion of a size
distribution on a log-log scale, and then applying a linear regression
fit to the logarithmic histogram in this ``scale-free'' range. While such 
measurements with a clearly visible power law range $[x_{p1},x_{p2}]$ are
still reasonably accurate, the problem becomes ill-defined when the
power law slope gradually varies from a flat segment at the lower end [$x_1$]
to a steepening edge at the upper end $[x_2]$) (for examples see Fig.~6.1 and
6.2 in Clauset et al.~2009). In principle, one could quantify the problem
by subdividing a size distribution into three zones, bound by the values
$x_1 \le x_{p1} \le x_{p2} \le x_2$, where the middle part $[x_{p1},x_{p2}]$
is fitted with a power law function, while the lower $[x_1, x_{p1}]$
and upper zone $[x_{p2}, x_2]$ are fitted with appropriate other model 
functions (as visualized in Fig.~2, right panel). 
However, such a method would still fail for data sets with
small-number statistics, because the pure power law range $[x_{p1}, x_{p2}]$
shrinks to arbitrary small ranges (see examples of flare events for single
stars (Figs.~14-15; second and third row). 

The third problem, the choice of the model, is discussed in the next 
section, where we argue which most likely effects should be included in 
models of power law-like distribution functions.

\subsection{		Evaluating Thresholded Power Laws		}

The ideal solution of model fitting is always a complete model that includes
all measurements or data points, quantified with a minimum number of free
parameters (Occam's razor criterion). For our size distribution fitting problem, 
this means that a complete model should describe not only the power law part
in the range $[x_{p1}, x_{p2}]$, but also the flattening in the lower portion
$[x_1, x_{p1}]$ and the steepening in the upper portion $[x_{p2}, x_2]$. 
We accomplish this goal simply by adding a threshold $x_0$ to the pure
power law function, i.e., $N(x) dx = (x + x_0)^{-a}$ (Eq.~5), which serves
a three-fold purpose: (1) $x_0$ quantifies a lower threshold above which
a data set is completely sampled; (2) $x_0$ could represent also a critical
physical threshold of an instability (in terms of an exponentially
growing avalanche); and (3) $x_0=-b_0$
can also be used to subtract an event-unrelated background $b_0$ from the
data before fitting of a model function is applied. Furthermore, the
largest event $x_2$ of the data set represents a sharp cutoff in the
differential size distribution, which produces a gradual steepening
in the cumulative size distribution (Eq.~10), which has to drop to zero
for values $x > x_2$. Fitting the correctly integrated cumulative
size distribution (Eq.~10) will then take care of the steepening near
the upper end of the observed size distribution.

How successful is this new method of thresholded power laws? First of all,
this method involves only one more free parameter ($x_0$) than fitting a 
straight power law distribution function (with two variables $n_0$ and $a$),
and thus it follows Occam's razor criterion. If we look at the 9 empirical
data sets in Fig.~9, we find that a thresholded power law is consistent
with the data ($\chi^2 \le 1.6$) in 4 out of the 9 cases, for which it
yields accurate power law slopes. In contrast, two cases are clearly not 
consistent (cities and fires), which probably could be fitted with a broken
power law, indicating two different scaling regimes. Regarding the
interpretation of these 9 empirical data sets, we can certainly consider
solar (flares) and geophysical phenomena (earthquakes and forest fires) 
as thresholded instabilities, as it was proposed for self-organized
criticality phenomena, while phenomena in sociophysics (city growth, 
power blackouts, terrorism attacks, stockmarket bubbles) have the 
statistics of extreme events in common, and information-type phenomena 
(words, surnames, weblinks) have the multiplicative behavior of combinational 
processes in common (such as in branching theory or percolation processes).

For solar flare data
the thresholded power law model is consistent with all data sets (HXRBS,
BATSE, RHESSI) regarding the peak count rate and total counts (Fig.~11),
while the size distributions for flare durations show some slight
deviations, which could indicate a bimodality between short and
long-duration flares. This is not surprising, because the duration
of long-lasting flares (several hours) is less well-defined due to
interruptions from the spacecraft orbit, and due to confusion of 
time-overlapping flare events. 

For stellar flare data we find consistency ($\chi^2 \lapprox 1$) of the 
thresholded power law model for all size distributions, for stellar flares
that are combined from many stars (Figs.~14-15; top row), as well as
for flare events from single stars (Figs.~14-15; second and third row).
This agreement is partially helped by the unavoidable small-number 
statistics, but it holds up to samples with 1538 flare events observed
with Kepler (Fig.~14-15; top row).

In summary, the thresholded power law seems to fit almost every data set,
and thus does not justify additional free model parameters.
The only two clear-cut deviations from a power law model are found for the
cases of city sizes and forest fires, which may indeed reveal multiple
scaling ranges, similar to multi-fractals in the geometric domain
(Kelty-Stephen et al.~2013).

\subsection{	Differential Versus Cumulative Distributions 	}

In this study we applied the thresholded power law model to both the
differential and the cumulative size distribution functions. What do we
learn from comparing these two methods? In principle we expect a higher
accuracy for cumulative size distribution functions, because they always
contain more datapoints per bin than the differential size distribution by 
definition. One difference is that the number of datapoints per bin
is independent in the differential size distribution, while it is not
for cumulative distributions. Nevertheless, our Monte-Carlo simulations
have proven that a goodness-of-fit criterion can be defined for both
methods (Eqs.~18-19), which both have a mean value of $\chi^2 \approx 1$
for random noise (although the value $\chi^2 \approx 0.8$ is slightly 
lower for cumulative distributions, see Fig.~5d). Applying both methods to 
the data, we find that the goodness-of-fit has the tendency to produce
better fits for the cumulative size distributions than for the differential 
size distributions. 

An important test is whether the two methods yield consistent values
of power law slopes. We show the power law slope values 
$\alpha \pm \sigma_\alpha$ with
errors in Fig.~16, which exhibits an excellent agreement between the
differential (Fig.~16, x-axis) and the cumulative distribution values
(Fig.~16, y=axis). Of course, the actual slope $b$ of a cumulative
distribution function is flatter by one, i.e., $b=a-1$. The consistency
of the two methods is largely obtained because both methods use the
same threshold value $x_0$ and thus an identical subset of events for
the fits. If the threshold $x_0$ is determined independently with the
two methods, a discrepancy in the resulting power law slope values $a$
would be unavoidable. In our study it was necessary to use both
fitting procedures simultaneously, because the differential size
distribution defines a more reliable threshold value $x_0$ (Fig.~6),
while the cumulative size distribution appears to give better fits
to the thresholded power law function (with a lower goodness-of-fit value
$\chi^2$).

\subsection{	Power laws and Physical Scaling Laws 		}

Power laws are ubiquous in statistical distributions of nonlinear energy
dissipation events, a scientific research area that started with the
model of self-organized criticality by Bak, Tang, and Wiesenfeld (1987).
We outlined in Section 2.2 a basic derivation of a power law distribution 
that can be generated by any nonlinear process that produces exponentially 
growing avalanches and is stopped after a random time interval. 
Reed and Hughes (2002) describe a class of generative models
for Pareto-type distributions that are multiplicative stochastic
processes that are killed (i.e., stopped or observed) by a secondary
process before equilibrium is established.
An early application of this model is the
accleration of cosmic ray particles (Fermi 1949), while later
applications considered accelerated particles, magnetic reconnection,
and nonpotential energy build-up in solar flares as 
avalanching processes (Rosner and Vaiana 1978; Aschwanden et al.~1998;
Lu and Hamilton 1991; Lu et al.~1993). Later applications included
not only solar and stellar flares, but also auroral emission and geomagnetic
substorms from planetary magnetospheres, particle events from
radiation belts, pulsar glitches, soft gamma-ray repeaters, blazars,
black-hole objects, cosmic rays, to boson clouds (for a recent review
see Aschwanden et al.~2015 and references therein).

A discussion of power laws in all astrophysical applications is beyond 
the scope of this study, but we just want to point out the most
fundamental reason why power laws are related to physical scaling laws.
A mathematical derivation of power law size distributions for various
observable parameters has been given in terms of the fractal-diffusive
self-organized criticality (FD-SOC) model (Aschwanden 2012, 2015).
The fundamental tenet of the FD-SOC model is the {\sl scale-free
probability conjecture}, which quantifies the statistical likelihood
of avalanche sizes $L$ with a reciprocal relationship to the volumes,
\begin{equation}
	N(L) dL \propto L^{-d} dL \qquad {\rm for}\ L \le L_{max} \ ,
\end{equation}
where $d$ is the Euclidean dimension and $L_{max}$ is the maximum
avalanche size or finite system size. Geometric relationships, such
as for the (avalanche) volume, $V \propto L^d$, predict then a
power law for the size distribution of volumes $V$,
\begin{equation}
        N(V) dV \propto N(L[V]) \left| {dL \over dV} \right| dV
                \propto V^{-(2-1/d)} dV \
                \propto V^{-\alpha_V} dV \ .
\end{equation}
Thus, in 3D space ($d=3$), we expect that the avalanche volumes
exhibit a size distribution of $N(V) \propto V^{5/3}$, with a power law
index $\alpha_V \approx 1.67$. If there is a nonlinear scaling
between the avalanche volume $V$ and an observable, such as the energy $E$,
which we quantify with the power index $\gamma$,
\begin{equation}
	E \propto V^\gamma \ , 
\end{equation}
we expect then a power law distribution $N(E)$ of  
\begin{equation}
        N(E) dE \propto N(V[E]) \left| {dV \over dE} \right| dE
                \propto E^{-(1-1/d)/\gamma} dE 
                \propto E^{-\alpha_E} dE \ .
\end{equation}
Based on this scaling of the power law index $\alpha_E=1+(1-1/d)/\gamma$
we can then directly evaluate the nonlinear scaling index $\gamma$ from
a measurement of the power law slope $\alpha_E$,
\begin{equation}	
	\gamma = {(1 - 1/d) \over (1 - \alpha_E)} \ .
\end{equation}
This analytical derivation of power law distributions is entirely based 
on statistical probabilities and observables that are connected with
geometric parameters by some power index. This explains the ubiquity
of power laws for the observables, but it does not explain the physics
of the underlying process. The physics comes in with the scaling law
between the observable and a geometric scale, i.e., $E \propto V^\gamma$.  
For a discussion of physical scaling laws applied to observed
astrophysical processes see a recent review (e.g., Aschwanden et al. 2015).

The introduction of a threshold value $x_0$ into a pure power law function
does not change the basic SOC concept, although it changes the value of
the power law slope at the lower and upper boundary of an observed
size distribution. Inclusion of this threshold value in fitting,
however, allows us to evaluate a more accurate value of the underlying
power law slope $\alpha$, such as the observed slope $\alpha_E$ for energy
size distributions (Eq.~25), and is therefore crucial for the 
identification or interpretation of the underlying physical scaling law. 
Since we have proven in this work that we can measure power law indices
of observed size distributions down to an accuracy of $\lapprox 1\%$,
we may refute the claim of Stumpf and Porter (2012) that the data
are insufficient for the identification of power law behavior to be
scientifically useful. 

\section{	SUMMARY AND CONCLUSIONS 		}

The fitting of ideal power law functions to real-world data is often
hampered by two major problems, the problem of insufficient statistics
(for extreme and rare events), and the problem of defining the correct
size range where a power law fit is applied. Deviations from ideal
power laws have been noted at the lower end, where often a smooth
rollover occurs due to incomplete sampling at the sensitivity limit,
and at the upper end, where a steepening in the cumulative size
distribution occurs due to the cutoff of the largest event.

In this study we modify the ideal power law function by adding a
constant $x_0$, which has a three-fold meaning; 
(1) as a physical threshold of an instability (or exponentially
growing avalanche process); and (2) as a sensitivity threshold 
for complete sampling; and (3) as a correction term for data
samples that are contaminated by event-unrelated background noise.
In addition we define an upper bound $x_2$ by the largest event, which
represents a sharp cutoff in the differential size distribution and
causes a gradual steepening at the upper end of the cumulative size
distribution. With the introduction of these modifications we can
fit most of the observed size distributions, as we tested with
9 mostly terrestrial data sets, 9 solar
flare data sets (from the HXRBS, BATSE, RHESSI missions), and 9 stellar 
flare data sets (from the Kepler mission). We like to point out that
the ``thresholded power law model'' used here is motivated by physical
arguments (instability threshold, detection threshold, background noise), 
while alternative mathematical functions, such as power laws with 
an exponential cutoff, exponential functions, stretched exponentials, 
or log-normal functions, which all have been used in fitting of 
empirical data (Clauset et al.~2009), may have been chosen for 
mathematical convenience, rather than based on a specific physical model.

Our recommended procedure to fit power laws in data sets consists of
the following four steps: (1) Histogram binning of both the differential 
and cumulative size distributions over the entire data range 
$[x_1, x_2]$; (2) Determination of a threshold $x_0$ from the bin with 
the maximum number of events in the differential size distribution; 
(3) Elimination of event-unrelated background $b_0$ by adjusting 
broken power laws to a straight power law in the fitting range $[x_0,x_2]$;  
and (4) Least-square fit by varying the power law slope $a$ and
normalization constant $n_0$ in the fitting range $[x_0,x_2]$,
with evaluation of the goodness-of-fit criterion that reveals
whether the thresholded power law model is consistent with the data
or not.

The practical value of our thresholded power law model is reflected
in a number of applications. Extreme events of solar flares, 
space weather storms, earthquakes, forest fires, blackouts,
and terrorism are of high interest for forecasting and mitigation 
of human risks and damage of electronic assets. Statistics of extreme 
events of stellar flares, cosmic rays, pulsar glitches, and gamma-ray
bursts allow us to test physical models operating under most violent
conditions in astrophysical particle accelerators and high-temperature 
plasmas. Measuring the power law slopes of multiple observables in an
avalanching system, possibly governed by self-organized criticality,
reveal us the degree of consistency with different models of physical 
scaling laws. The evaluation of threshold values is important for
understanding the onset of instabilities and their triggers. 
In summary, there is a large number of applications in complexity
theory and self-organized criticality models that benefit from
reliable measurements of power law indices.

\acknowledgments
The author acknowledges the hospitality and partial support for two
workshops on ``Self-Organized Criticality and Turbulence'' at the
{\sl International Space Science Institute (ISSI)} at Bern, Switzerland,
during October 15-19, 2012, and September 16-20, 2013, as well as
constructive and stimulating discussions with 
Sandra Chapman, 
Paul Charbonneau, 
Aaron Clauset, 
Michaila Dimitropoulou,
Manolis Georgoulis,
Stefan Hergarten, 
Henrik Jeldtoft Jensen,
James McAteer, 
Shin Mineshige,
Laura Morales,
Mark Newman,
Naoto Nishizuka,
Gunnar Pruessner, 
John Rundle, 
Surja Sharma,
Antoine Strugarek,
Vadim Uritsky,
and Nick Watkins.
We thank also the anonymous statistical expert who provided links
to the statistical literature, and the anonymous referee for thoughtful
comments. This work was partially supported by NASA contract NNX11A099G
``Self-organized criticality in solar physics'' and NASA contract
NNG04EA00C of the SDO/AIA instrument to LMSAL. 

\clearpage
                %REFERENCES
\section*{	References				}

\def\ref#1{\par\noindent\hangindent1cm {#1}}

\ref{Arnold, B.C. 2015 (2nd edition), 1983 (1st edition), {\sl ``Pareto 
	Distributions''}, International Co-operative Publishing House.
	ISBN 0-89974-012-X.} 
\ref{Aschwanden, M.J., Dennis,B.R., and Benz,A.O. 
	1998, ApJ 497, 972.}
\ref{Aschwanden, M.J. 2004, {\sl Physics of the Solar Corona - 
	An Introduction} (1st Edition), Praxis Publishing Ltd., 
	Chichester UK, and Springer, New York, ISBN 3-540-22321-5, 
	(842pp).}
\ref{Aschwanden, M.J. 2011a, Solar Phys. 274, 99.}
\ref{Aschwanden, M.J. 2011b, SP 274, 119.}
\ref{Aschwanden, M.J. 
	2011c, {\sl Self-Organized Criticality in Astrophysics.
	The Statistics of Nonlinear Processes in the Universe}, 
	Springer and PraXis: Berlin.}
\ref{Aschwanden, M.J. 
	2012, AA 539, A2 (15 p).}
\ref{Aschwanden, M.J., Crosby, N., Dimitropoulou, M., Georgoulis, M.K., Hergarten, S., 
	McAteer, J., Milovanov, A., Mineshige, S., Morales, L., Nishizuka, N., 
	Pruessner, G., Sanchez, R., Sharma, S., Strugarek, A., and Uritsky, V.
	2015, Space Science Reviews (published online first), 
	DOI 10.1007/s11214-014-0054-6,
 	{\sl 25 Years of Self-Organized Criticality: Solar and Astrophysics}}
\ref{Bak, P., Tang, C., and Wiesenfeld,K. 1987, PhRvL 69/4, 381.}
\ref{Balona, L.A. 2015, MNRAS 447, 2714.}
\ref{Bauke, H. (2007), Eur. Phys. J. B, 58, 167.}
\ref{Baumjohann, W. and Treumann, R.A. 1996, {\sl Basic space plasma physics},
 	Imperial College Press, London.}
\ref{Begelman, M.C., Volonteri, M., and Rees, M.J. 2006, MNRAS 370, 289.}
\ref{Bellan, P.M. 2006, {\sl Fundamentals of plasma physics},
	Cambridge University Press, Cambridge.}
\ref{Benz, A.O. 1993, {\sl Plasma astrophysics, kinetic processes in solar 
	and stellar coronae}, Kluwer Academic Publishers, Dordrecht, 
	The Netherlands.}
\ref{Biesecker, D.A., Ryan, J.M., Fishman, G.J. 1993,
        Lecture Notes in Physics 432, 225.}
\ref{Biesecker, D.A. 1994,
        {\sl On the occurrence of solar flares observed with the
        Burst and Transient Source Experiment (BATSE)},
        PhD Thesis, University of New Hampshire.}
\ref{Biskamp, D. 2000, {\sl Magnetic reconnection in plasmas},
	Cambridge: Cambridge University Press.}
\ref{Boss, A.P. 1997, Science 276,(5320), 1836.}
\ref{Broder, A., Kumar, R., Maghoul, F., Raghavan, P., Rajagopalan, S.,
	Stata, R.,  Tomkins, A., and Wiener, J., 2000, Computer Networks 33, 309.}
\ref{Bromund, K.R., McTiernan, J.M., Kane, S.R. 1995,
        ApJ 455, 733.}
\ref{Christe, S., Hannah, I.G., Krucker, S., McTiernan, J., and Lin, R.P.
        2008, ApJ 677, 1385.}
\ref{Clauset, A., Young, M., and Gleditsch, K.S. 2007, J. Conflict Resolution, 51, 58.} 
\ref{Clauset, A., Shalizi, C.R., and Newman, M.E.J. 2009, SIAM Rev. 51/4, 661.}
\ref{Cowling, T.G. 1976, {\sl Magnetohydrodynamics}, Monographs on Astronomical
	Subjects 2, Adam Hilger Ltd., Bristol.}
\ref{Crosby, N.B., Aschwanden, M.J., and Dennis, B.R. 1993,
        Solar Phys. 143, 275.}
\ref{Crosby, N.B. 1996, {\sl Contribution \`a l'Etude des Ph\'enom\`enes Eruptifs
        du Soleil en Rayons \`a partir des Observations de l'Exp\'erience
        WATCH sur le Satellite GRANAT}, PhD Thesis, Universit\'e Paris VII, Meudon, Paris.}
\ref{D'Agostino, R.B., and Stephens, M.A. (eds.), 1986, {\sl Goodness-of-fit Techniques},
	Marcel Dekker, New York.}
\ref{Datlowe, D.W., Elcan, M.J., and Hudson, H.S. 1974, Solar Phys. 39, 155.}
\ref{Dennis, B.R. 1985, SP 100, 465.}
\ref{Feigenbaum, J. 2003, Rep. Prog. Phys. 66, 1611.}
\ref{Fermi, E. 1949, Phys. Rev. Lett. 75, 1169.}
\ref{Georgoulis, M.K., Vilmer,N., and Crosby,N.B. 2001, AA 367, 326.}
\ref{Galam, S. 2012, {\sl Sociophysics. A physicist's modeling of psycho-political
	phenomena}, Springer: New York.}
\ref{Giles, D.E., Feng, H., and Godwin, R. 2011, Econometrics Working Paper
	EWP1104, ISSN 1485-6441.}
\ref{Goldstein, M.L., Morris, S.A., and Yen, G.G. 2004, Eur. Phys. J. B, 41, 255.}
\ref{Goossens, M. 2003, ApSS Library Vol. 294, Dordrecht: Kluwer Academic Publishers,
	(405pp).}
\ref{Hosking, J.R.M. and Wallis, J.R. 1987, Technometrics 29(3), 339.}
\ref{Huberman, B.A. and Adamic, L. 1999, Nature 401, 131.}
\ref{Johnson, N.L., Kotz, S., Balakrishnan, N. 1994, {\sl Continuous
	univariate distributions}, Vol. 1, Wiley Series in Probability and
	Statistics.}
\ref{Katz, J. 1986, JGR 91, 10412.}
\ref{Kelty-Stephen, D.G., Palatinus, K., Saltzman, E., and Dixon, J.A.
	2013, Ecological Psychology 25, 1.}
\ref{Kivelson, M.G. and Russell, C.T. 1995, {\sl Introduction to space physics},
 	Cambridge University Press, Cambridge, 568p.}
\ref{Koch, D.G., Borucki, W.J., Basrie, G., et al. 2010, ApJ 713, 79.}
\ref{Krall, K.R. and Trivelpeace 1971, {\sl Principles of Plasma Physics},
	McGraw-Hill Book Company.}
\ref{Lee, T.T., Petrosian, V., and McTiernan, J.M. 1993,
        ApJ 412, 401.}
\ref{Lin, R.P., Schwartz, R.A., Kane, S.R., Pelling, R.M., Hurley, K.C.
        1984, ApJ 283, 421.}
\ref{Lin, R.P., Feffer,P.T., and Schwartz,R.A. 2001, ApJ 557, L125.}
\ref{Lomax, K.S. 1954, J. American Statistical Association 49, 847.}
\ref{Lu, E.T. and Hamilton, R.J., 1991, ApJ  380, L89.}
\ref{Lu, E.T., Hamilton, R.J., McTiernan, J.M., and Bromund, K.R. 1993, ApJ 412, 841.}
\ref{Lui, A.T.Y. 1991, JGR 96, 1849.}
\ref{Maehara, H., Shibayama, T., Notsu, S. et al.~2012, Nature 485, 478.}
\ref{Mandelbrot, B.B. 1977,
        {\sl Fractals: form, chance, and dimension}, Translation of
        {\sl Les objects fractals}, W.H. Freeman, San Francisco.}
\ref{Melrose, D.B. 1980a, {\sl Plasma Astrophysics. Nonthermal Processes 
	in Diffuse Magnetized Plasmas. Volume 1: The Emission, Absorption 
	and Transfer of Waves in Plasmas},
 	Gordon and Breach Science Publishers, New York.}
\ref{Melrose, D.B. 1980b, {\sl Plasma Astrophysics. Nonthermal Processes 
	in Diffuse Magnetized Plasmas. Volume 2: Astrophysical Applications},
 	Gordon and Breach Science Publishers, New York.}
\ref{Melrose, D.B. 1986, {\sl Instabilities in space and laboratory plasmas},
	Cambridge University Press, Cambridge.}
\ref{Newman, M.E.J. 2005, Contemporary Physics 46, 323.}
\ref{Perez Enriquez, R., Miroshnichenko, L.I. 1999, Sol. Phys. 188, 169.}
\ref{Press, W.H., Flannery, B.P., Teukolsky, S.A., and Vetterling, W.T.
 	1986, {\sl Numerical recipes, The Art of Scientific Computing}, Cambridge.}
\ref{Priest, E.R. 1982, {\sl Solar Magnetohydrodynamics}, Geophysics and Astrophysics
	Monographs, Vol. 21, D.Reidel Publishing Company, Dordrecht.}
\ref{Priest, E.R. and Forbes, T. 2000, {\sl Magnetic reconnection (MHD Theory and
	Applications)}, Cambridge: Cambridge University Press.}
\ref{Priest, E.R. 1978, Solar Phys. 58, 57.}
\ref{Pruessner, G. 2012, {\sl Self-organised criticality. Theory, models 
	and characterisation}, Cambridge University Press.}
\ref{Reed, W.J. and Hughes, B.D. 2002, Phys. Rev. Lett. E 66, 067103.}
\ref{Rosner, R. and Vaiana, G.S. 1978 ApJ 222, 1104.}
\ref{Schmidt, G. 1979, {\sl Physics of high temperature plasmas}, (2nd edition),
	Academic Press, New York.}
\ref{Schwartz, R.A., Dennis, B.R., Fishman, G.J., Meegan, C.A., Wilson, R.B.,
        Paciesas, W.S. 1992, in Shrader, C.R., Gehrels, N., Dennis, B.R. (eds.),
        {\sl The Compton Observatory Science Workshop}, NASA CP {\bf 3137}, 
	NASA: Washington DC, 457.}
\ref{Shibayama, T., Maehara, H., Notsu, S. et al.~2013, ApJSS 209, 5.}
\ref{Shibaskai, K. 2001, ApJ 557, 326.}
\ref{Smith, R.A., Goldstein, M.L., and Papadopoulos, K. 1979, ApJ 234, 348.}
\ref{Somov, B.V. 2000, {\sl Cosmic plasma physics}, ApSS Library, Vol. 251, (672pp.),
	Kluwer Academic Publishers, Dordrecht.}
\ref{Springel, V., White, S.D.M., Jenkins, A., et al. 2005, Nature 435, 629.}
\ref{Stumpf, M.P.H. and Porter, M.A. 2012, Science 335, issue 10 Feb 2012.} 
\ref{Sturrock, P.A. 1994, {\sl Plasma physics. An introduction to the theory 
	of astrophysica, geophysical and laboratory plasmas},
 	Cambridge University Press, Cambridge.}
\ref{Su, Y., Gan, W.Q., and Li, Y.P. 2006, Solar Phys. 238, 61.}
\ref{Tajima, T. and Shibata, K. 2002, {\sl Plasma astrophysics},
	Perseus Publishing, Cambridge, Massachusetts.}
\ref{Tranquille, C., Hurley, K., and Hudson, H.S. 2009, Solar Phys. 258, 141.}
\ref{Treumann, R.A. and Baumjohann, W. 1997, {\sl Advanced space plasma physics},
 	Imperial College London.}
\ref{Turcotte, D.L. 1999, Phys. Earth Planet Inter. 111, 275.}
\ref{Walkowicz, L.M., Basri, G., Batalha, N. et al. 2011, ApJ 141, 50.}
\ref{Wang, B. and Silk, J. 1994, ApJ 427, 759.}
\ref{Willis, J.C. and Yule, G.U. 1922, Nature 109, 177.}
\ref{Wu, C.J., Ip, W.H., and Huang, L.C. 2015, ApJ 798, 92.}
\ref{Wu, C.S. and Lee, L.C. 1979, ApJ 230, 621.}

\clearpage

%_______________TABLE 1 _____________________________________

\begin{table}
\begin{center}
\normalsize
\caption{A selection of hydrodynamic, magneto-hydrodnamic, and kinetic instabilities 
and threshold conditions, occurring in solar and astrophysical plasmas.}
\begin{tabular}{ll}
\hline
\hline
Instability                                             &Threshold condition     	\\
\hline
{\bf Interchange or Pressure-Driven Instabilities:}	&                       	\\
$\quad${\sl Rayleigh$-$Taylor instability:}        	&                       	\\
$\qquad$ {\sl Hydrodynamic}                      	&${\bf g} \cdot \nabla n_0 < 0$ \\
$\qquad$ {\sl Hydromagnetic (Kruskal$-$Schwarzschild)}	&${\bf k} \cdot {\bf B} = 0$    \\
$\qquad$ {\sl Hydromagnetic (Parker instability)}	&${\bf k} \cdot {\bf B} \neq 0$ \\
$\quad${\sl Kelvin$-$Helmholtz instability}       	&                       	\\
$\qquad$ {\sl Hydromagnetic:}                     	&$v_1 > v_{A,2}$        	\\
$\quad${\sl Ballooning instability}               	&${\bf j} \times {\bf B} > \rho {\bf g}$ \\
                                                        &                       	\\
{\bf Thermal Instabilities:}                         	&                       	\\
$\quad${\sl Convective instabilities}             	&$(dT/dz)_{crit}$ 		\\
$\quad${\sl Radiatively-driven thermal instabilities}	&${\tau}_{cond} > {\tau}_{rad}$ \\ 
$\quad${\sl Heating-driven thermal instabilities} 	&$s_H/L < 1/3$           	\\
                                                        &                       	\\
{\bf Resistive Instabilities:}                       	&                       	\\
$\quad${\sl Gravitational mode}                   	&$F_{grav} > ({\bf j} \times {\bf B})$\\
$\quad${\sl Rippling mode}                        	&$F_{adv} > ({\bf j} \times {\bf B})$\\
$\quad${\sl Tearing mode}                         	&$(dB/dx)_{crit}$       	\\
                                                        &                       	\\
{\bf Current Pinch Instabilities:}                   	&                       	\\
$\quad${\sl Cylindrical pinch:}                    	&                       	\\
$\qquad${\sl Kink mode}                          	&$B_{0\varphi}^2 \ln(L/a) > B_{0z}^2$ \\
$\qquad${\sl Sausage mode}                       	&$B_{0\varphi}^2 > 2 B_{0z}^2$ 	\\
$\qquad${\sl Helical/torsional mode}             	&$B_{0\varphi} > (2 \pi a/L) B_{0z}$ \\
$\quad${\sl Current sheet:}                        	&                       	\\
							&				\\
{\bf Kinetic Instabilities:}                   		&                       	\\
$\quad${\sl Bump-in-tail instability}                   &$df(v_\parallel)/dv_\parallel) \gapprox 0$\\ 
$\quad${\sl Loss-cone instability}                  	&$df(v_\perp)/dv_\perp) \gapprox 0$\\
$\quad${\sl Buneman instability}			&$v_d \gapprox 1.7 (v_{te}+v_{ti}$)\\
\hline
\end{tabular}
\end{center}
\end{table}

%_______________TABLE 2 _____________________________________
\begin{table}
\caption{Power law slopes of (terrestrial and empirical) data 
(Clauset et al. 2009).}
\medskip
\begin{tabular}{lrrrrrrrr}
\hline
Dataset         &Number & Threshold & Back-  & Differential  & Cumulative   & Clauset & Differential  & Cumulative\\
                &       &           & ground & slope         & slope        & slope   & fit & fit \\
                &       & $x_0$     & $b_0$  & $a_{diff}$    & $a_{cum}$    & $a$     & $\chi^2$ & $\chi^2$ \\
\hline
(a) Flares      & 12772 &  15.8	&    57	& 1.72$\pm$0.02 & 1.72$\pm$0.02 & 1.79$\pm$0.02 & 0.9  & 0.6 \\
(b) Quakes      & 19301 &   861 &     1	& 1.85$\pm$0.02 & 1.84$\pm$0.01 & 1.64$\pm$0.04 & 2.5  & 1.3 \\
(c) Fires       &203784 &  0.63	&     1	& 1.42$\pm$0.01 & 1.48$\pm$0.01 & 2.20$\pm$0.30 &23.2  &15.1 \\
(d) Cities      & 19446 &   631	&     0	& 1.82$\pm$0.02 & 1.85$\pm$0.01 & 2.38$\pm$0.08 & 5.2  & 7.2 \\
(e) Blackouts   &   210 & 39800	&     0	& 1.84$\pm$0.15 & 1.88$\pm$0.13 & 2.30$\pm$0.03 & 0.6  & 0.4 \\
(f) Terrorism   &  9100 &     1	&     2	& 1.99$\pm$0.04 & 2.00$\pm$0.04 & 2.40$\pm$0.02 & 4.0  & 2.7 \\
(g) Words       & 18855 &     1 &     1	& 1.98$\pm$0.02 & 1.92$\pm$0.02 & 1.95$\pm$0.02 & 7.0  & 3.2 \\
(h) Surnames    & 19301 & 19700	&     2	& 2.74$\pm$0.07 & 2.61$\pm$0.06 & 2.50$\pm$0.02 & 2.4  & 1.1 \\
(i) Weblinks    & 2.4 $10^8$ &1	&     1	& 2.35$\pm$0.00 & 2.13$\pm$0.00 & 2.34$\pm$0.01 &1078  & 623 \\
\hline
\end{tabular}
\end{table}

%_______________TABLE 2 _____________________________________
\begin{table}[t]
\caption{Frequency distributions measured from solar flares in hard X-rays
and $\gamma$-rays. The prediction is based on the FD-SOC model (Aschwanden 2012).}
\medskip
\begin{tabular}{lllrll}
\hline
Power law        &Power law       &Power law   &Number     &Instrument&References\\
slope of        &slope of       &slope of   &of         &and     &\\
peak flux       &fluence        &durations  &events     &threshold&\\
$\alpha_P$      &$\alpha_E$     &$\alpha_T$ &$n$        &energy  &\\
\hline
\hline
1.8             &               &           &123        &OSO--7($>$20 keV)&$^1)$Datlowe \etal (1974) \\
2.0             &               &           &25         &UCB($>$20 keV)   &$^2)$Lin \etal (1984)\\
1.8             &               &           &6775       &HXRBS($>$20 keV) &$^3)$Dennis (1985)\\
1.73$\pm$0.01   &               &           &12,500     &HXRBS($>$25 keV) &$^4)$Schwartz \etal (1992)\\
1.73$\pm$0.01   &1.53$\pm$0.02  &2.17$\pm$0.05 &7045    &HXRBS($>$25 keV) &$^5)$Crosby \etal (1993)\\
1.71$\pm$0.04   &1.51$\pm$0.04  &1.95$\pm$0.09 &1008    &HXRBS($>$25 keV) &$^6)$Crosby \etal (1993)\\
1.68$\pm$0.07   &1.48$\pm$0.02  &2.22$\pm$0.13 &545     &HXRBS($>$25 keV) &$^7)$Crosby \etal (1993)\\
1.67$\pm$0.03   &1.53$\pm$0.02  &1.99$\pm$0.06 &3874    &HXRBS($>$25 keV) &$^8)$Crosby \etal (1993)\\
1.61$\pm$0.03   &               &           &1263       &BATSE($>$25 keV) &$^9)$Schwartz \etal (1992)\\
1.75$\pm$0.02   &               &           &2156       &BATSE($>$25 keV) &$^{10}$Biesecker \etal (1993)\\
1.79$\pm$0.04   &               &           &1358       &BATSE($>$25 keV) &$^{11}$Biesecker \etal (1994)\\
1.59$\pm$0.02   &               &2.28$\pm$0.08 &1546    &WATCH($>$10 keV) &$^{12}$Crosby (1996)\\
1.86            &1.51           &1.88       &4356       &ISEE--3($>$25 keV) &$^{13}$Lu \etal (1993)\\
1.75            &1.62           &2.73       &4356       &ISEE--3($>$25 keV) &$^{14}$Lee \etal (1993)\\
1.86$\pm$0.01   &1.74$\pm$0.04  &2.40$\pm$0.04 &3468    &ISEE--3($>$25 keV) &$^{15}$Bromund \etal (1995)\\
1.80$\pm$0.01   &1.39$\pm$0.01  &           &110        &PHEBUS($>$100 keV) &$^{16}$Perez-Enriquez \& \\
                &               &           &           &                   &\quad Miroshnichenko (1999)\\
1.80$\pm$0.02   &               &2.2$\pm$1.4&2759       &RHESSI($>$12 keV)  &$^{17}$Su \etal (2006)\\
1.58$\pm$0.02   &1.7$\pm$0.1    &2.2$\pm$0.2&4241       &RHESSI($>$12 keV)  &$^{18}$Christe \etal (2008)\\
1.6             &               &           &243        &BATSE($>$8 keV)    &$^{19}$Lin \etal (2001)\\
1.61$\pm$0.04   &               &           &59         &ULYSSES($>$25 keV) &$^{20}$Tranquille \etal (2009)\\
\hline
1.74$\pm$0.02 	&1.66$\pm$0.02 	&2.54$\pm$0.02 &10291	&HXRBS($>$25 keV)   &$^{21}$This work (cumulative)\\
1.75$\pm$0.02 	&1.69$\pm$0.03 	&2.57$\pm$0.03 &10291	&HXRBS($>$25 keV)   &$^{22}$This work (differential)\\
1.79$\pm$0.02 	&1.61$\pm$0.03 	&2.32$\pm$0.03 & 7233	&BATSE($>$25 keV)   &$^{23}$This work (cumulative)\\
1.89$\pm$0.02 	&1.63$\pm$0.04 	&2.23$\pm$0.03 & 7233	&BATSE($>$25 keV)   &$^{24}$This work (differential)\\
1.81$\pm$0.03 	&1.62$\pm$0.02 	&1.96$\pm$0.02 &11549	&RHESSI($>$25 keV)  &$^{25}$This work (cumulative)\\
1.82$\pm$0.04 	&1.58$\pm$0.02 	&1.95$\pm$0.02 &11549	&RHESSI($>$25 keV)  &$^{26}$This work (differential)\\
\hline
{\bf 1.67}      &{\bf 1.50}     &{\bf 2.00} &           &FD-SOC prediction  &Aschwanden (2012)\\
\hline
\end{tabular}
\end{table}

%_______________TABLE 3 _____________________________________
\begin{table}
\caption{Frequency distributions of bolometric energies radiated in stellar flares observed with Kepler.}
\medskip
\begin{tabular}{llrlll}
\hline
Star            &Instrument     &Number          &Power law      &power law	&References: \\
                &               &of events       &slope          &slope  	&            \\
                &               &                &differential   &cumulative    &            \\
                &               &                &$\alpha_{dif}$ &$\alpha_{cum}$&            \\
\hline
G5-stars        &Kepler         &  365           & 2.3$\pm$0.3   &		 &Maehara et al.~(2012)  \\
G5-stars slow   &Kepler         &  101           & 2.0$\pm$0.2   &		 &Maehara et al.~(2012)  \\
G5-stars        &Kepler         & 1547           & 2.2           &		 &Shibayama et al.~(2013)\\
G5-stars slow   &Kepler         &  397           & 2.0           &		 &Shibayama et al.~(2013)\\
G5-stars        &Kepler         & 1538           & 2.04$\pm$0.13 &		 &Aschwanden (2015)      \\
G5-stars        &Kepler         & 1538           & 2.43$\pm$0.08 & 2.42$\pm$0.06 &This study  \\
G-type stars	&Kepler		& 4494           & 2.04$\pm$0.17 &		 &Wu et al. (2015) \\
K-M,A-F stars	&Kepler		&  209           & 1.69$\pm$0.16 & 1.71$\pm$0.12 &Balona (2015), This study\\
KID3557532	&Kepler		&  196           & 2.11$\pm$0.19 &		 &Wu et al. (2015) \\
KID6034120      &Kepler         &   45           & 3.12$\pm$0.60 & 3.17$\pm$0.48 &This study \\
KID6697041      &Kepler         &   37           & 1.83$\pm$0.37 & 1.51$\pm$0.25 &This study \\
KID6865416	&Kepler		&  147           & 1.77$\pm$0.10 &		 &Wu et al. (2015) \\
KID75264976	&Kepler		&   40           & 1.92$\pm$0.34 & 1.98$\pm$0.32 &This study \\
KID7532880	&Kepler		&  159           & 1.90$\pm$0.16 &		 &Wu et al. (2015) \\
KID8074287	&Kepler		&  160           & 1.87$\pm$0.10 &		 &Wu et al. (2015) \\
KID8479655 	&Kepler		&   39           & 1.45$\pm$0.23 & 1.47$\pm$0.24 &This study \\
KID8547383 	&Kepler		&   40           & 3.41$\pm$0.68 & 2.58$\pm$0.41 &This study \\
KID9653110	&Kepler		&  158           & 1.64$\pm$0.07 &		 &Wu et al. (2015) \\
KID10422252	&Kepler		&  177           & 1.75$\pm$0.08 &		 &Wu et al. (2015) \\
KID10422252     &Kepler         &   57           & 2.99$\pm$0.58 & 2.78$\pm$0.37 &This study \\
KID10745663	&Kepler		&  137           & 1.63$\pm$0.10 &		 &Wu et al. (2015) \\
KID11551430	&Kepler		&  202           & 1.59$\pm$0.06 &		 &Wu et al. (2015) \\
\hline
\end{tabular}
\end{table}

\clearpage
%__________________________FIGURE_________________________ 

\begin{figure}
\plotone{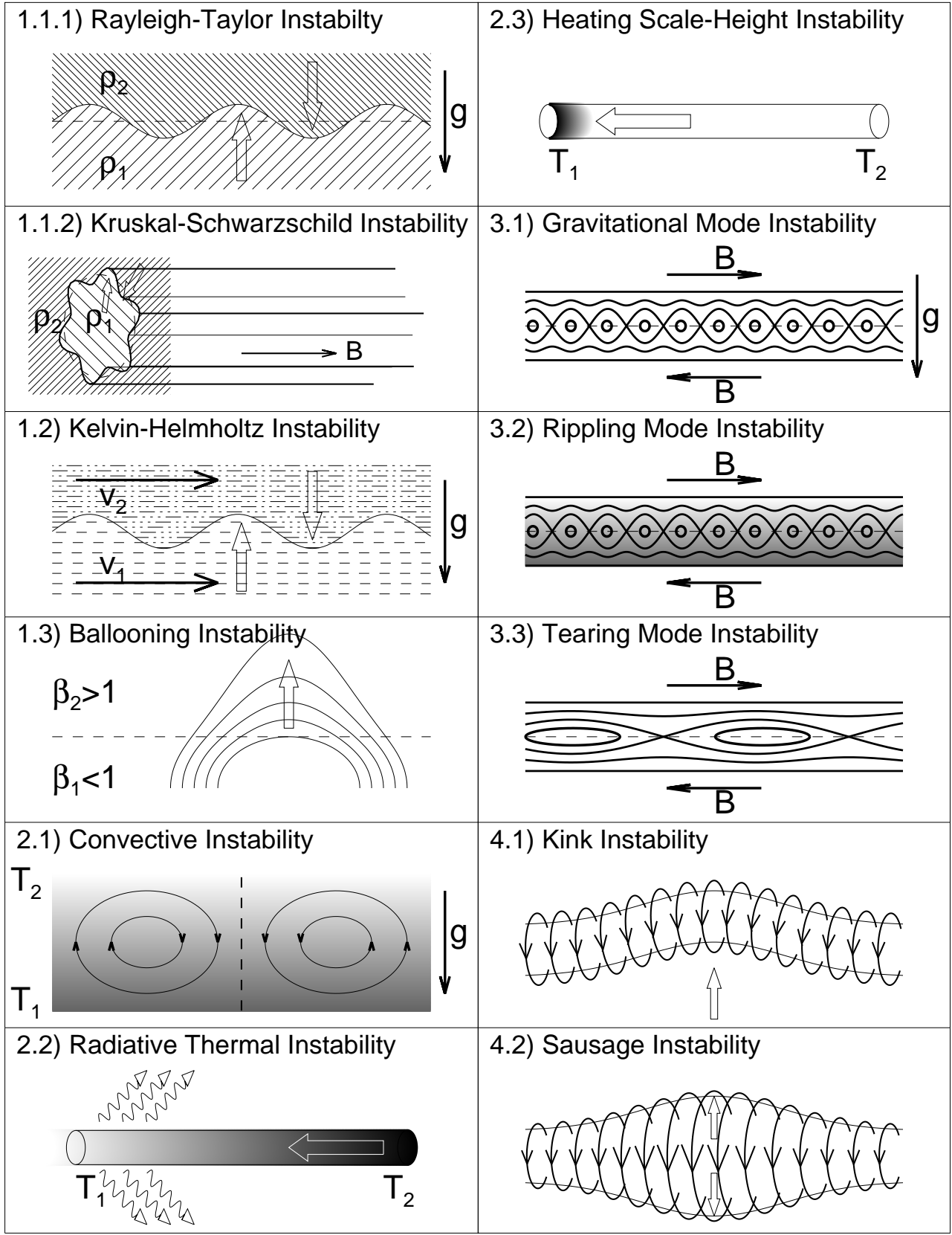}
\caption{Hydrodynamic and magneto-hydrodynamic instabilities that can occur in 
astrophysical plasmas are illustrated (in the same order as in Table 1). 
Different densities (${\rho}_1,
{\rho}_2$) are rendered with hatched linestyle, different velocities (v$_1$, v$_2$)
with dashed linestyle, temperature gradients ($T_1, T_2$) with greyscales,
longitudinal magnetic field lines ($B_0$) with thin solid lines,
azimuthal magnetic field components ($B_{\varphi}$) with thick solid
lines, and radiation with wiggly lines. The directions of the disturbances that
lead to an instability are indicated with thick white arrows (Aschwanden 2004).}
\end{figure}

\begin{figure}
\plotone{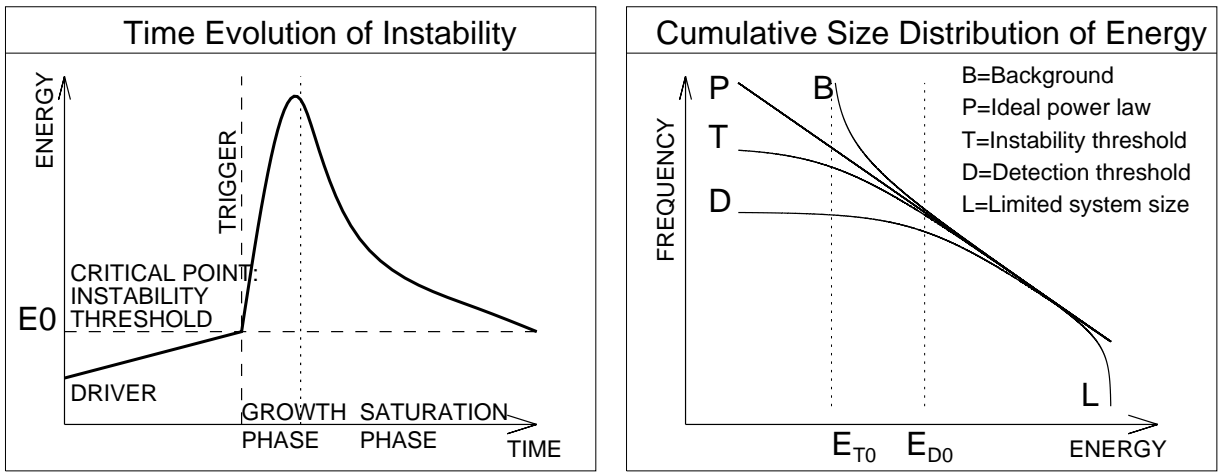}
\caption{{\sl Left:} Schematic time evolution of an instability, which has an onset
when the energy parameter exceeds a critical point or instability threshold $E_{T0}$.
{\sl Right:} Deviations from an ideal power law (P) of the cumulative size
distribution of energies occur when a background is present (B), when the sample
contains values below the instability threshold $E_{T0}$ (T), when the instrumental
sensitivity sets a detection threshold $E_{D0}$ (D), and when a (finite) limited system
size is present (L).}
\end{figure}

\begin{figure}
\plotone{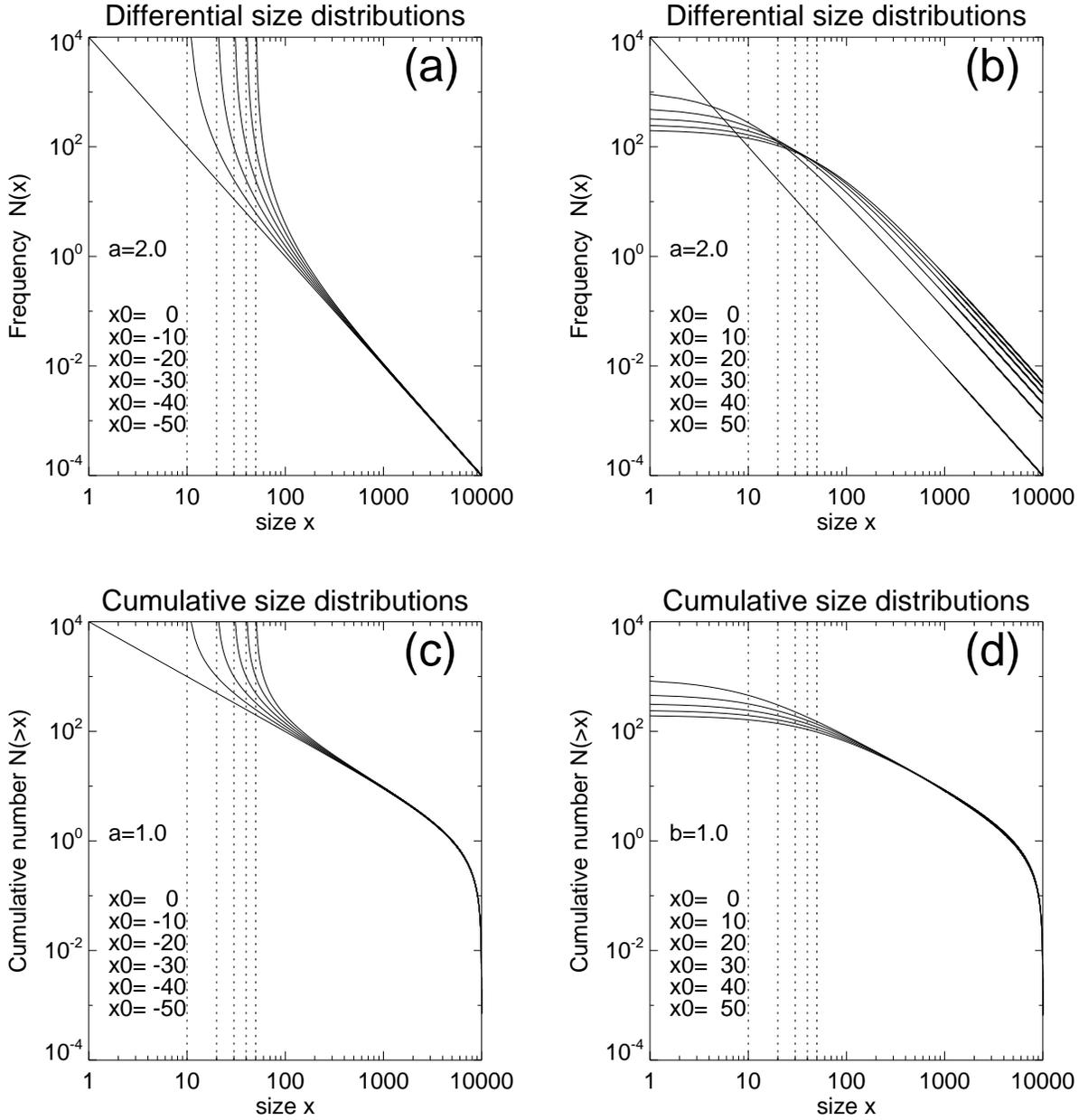}
\caption{Differential (top) and cumulative occurrence frequency 
distributions (bottom) for a power law slope of 
$a=2.0$ (or cumulative slopes $b=a-1=1.0$) and a range of
negative (left) and positive (right) threshold values $x_0$.
Note that the cumulative distributions show a steep drop-off 
at the maximum value near $x \lapprox 10^4$.}
\end{figure}

\begin{figure}
\plotone{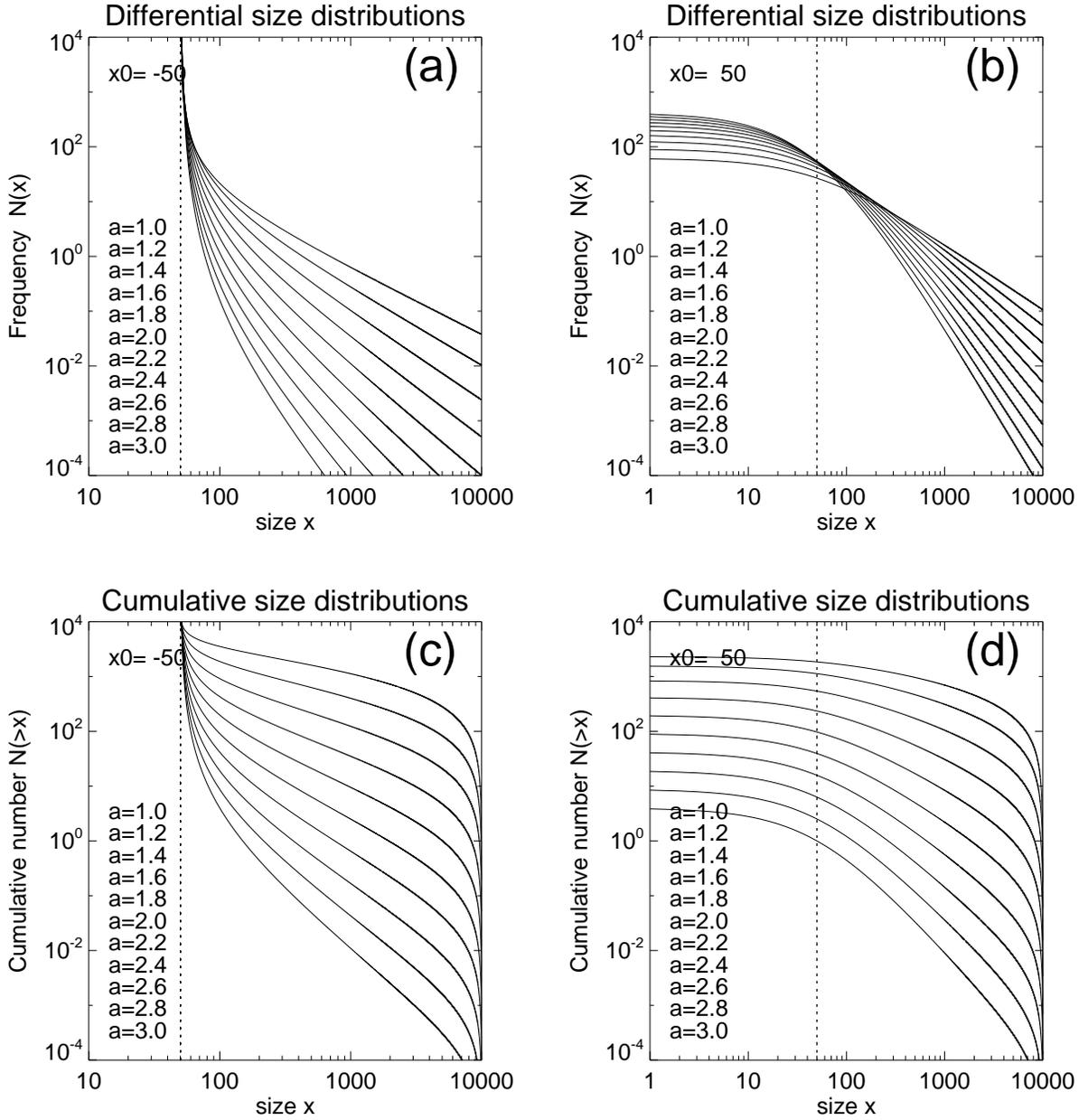}
\caption{Differential (top) and cumulative occurrence frequency 
distributions (bottom) for a range of power law slopes of 
$a=1.0,...,3.0$ (or cumulative slopes $b=a-1$), with a 
negative (left) or positive (right panels) threshold value $x_0$.
Note that the cumulative distributions show a steep drop-off 
at the maximum value near $x \lapprox 10^4$.}
\end{figure}

\begin{figure}
\plotone{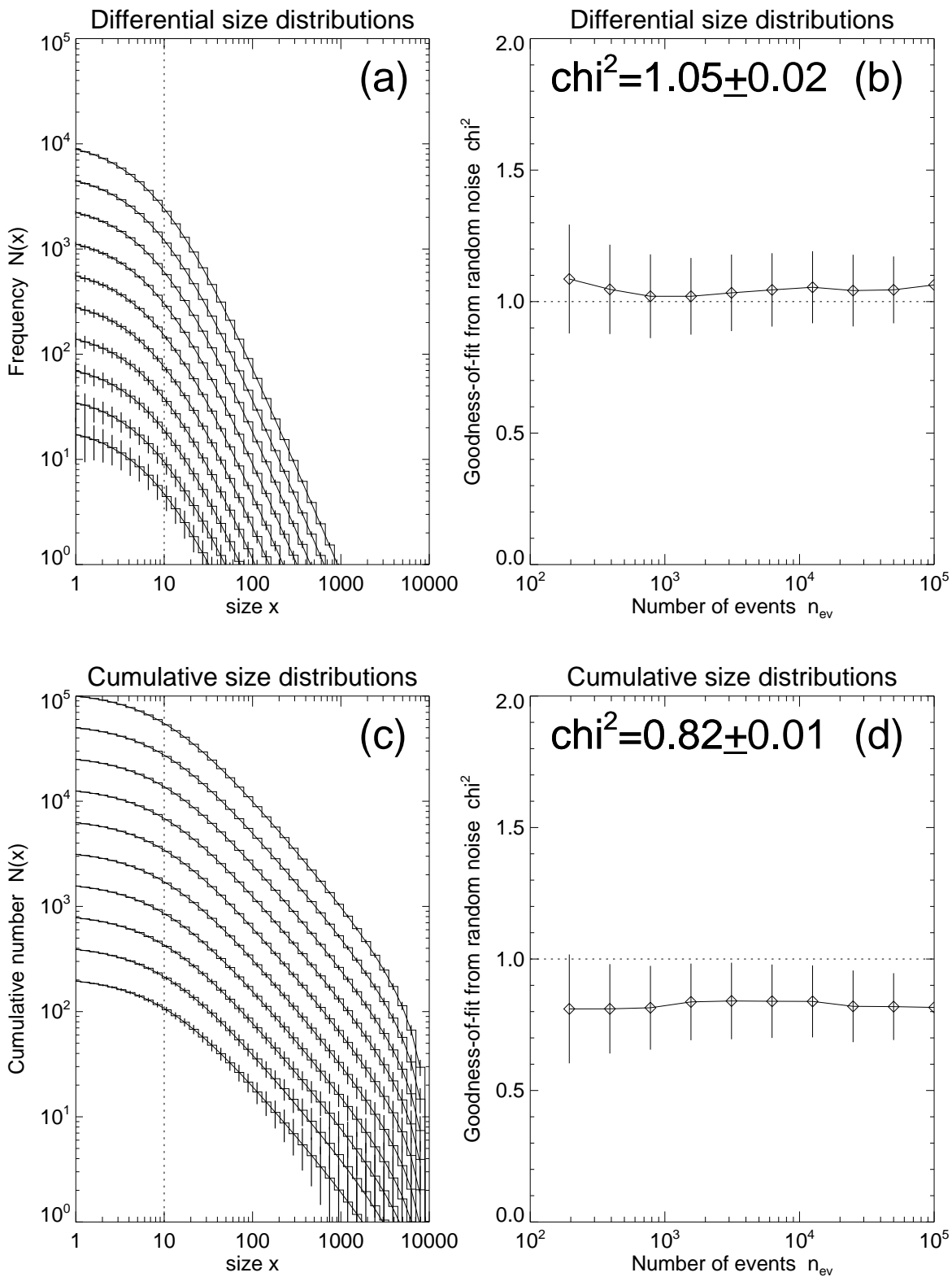}
\caption{Numerical simulations of differential (top panels) and 
cumulative (bottom panels) size distributions for thresholded
power law distributions with the parameters $x_0=10, a=2.0$, 
$n_{ev}=10^5,...,195$, where the number of events decreases by
a factor of two for each of the 10 data sets. The theoretical
distribution function is plotted with a smooth curve, while the
simulated data using a random generator are shown in form of
histograms with error bars, each set showing the average 
and standard deviations {\bf from 1000 different random number
realizations} (left panels). The resulting goodness-of-fit 
values $\chi^2$ are plotted as a function of the number of 
events per data set (right panels).}
\end{figure}

\begin{figure}
\plotone{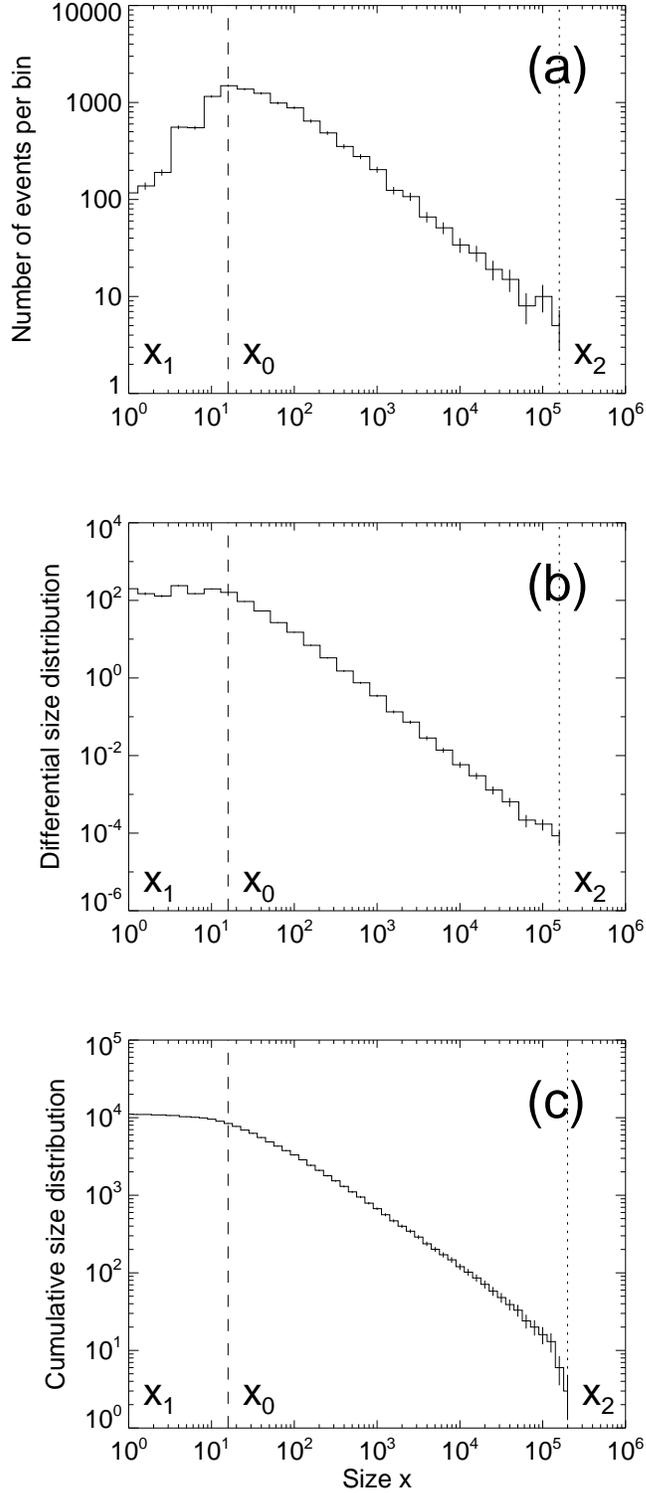}
\caption{The determination of the lower threshold $x_0$ of
a completely sampled power law distribution is obtained from
the maximum in the distribution of sampled events per bin
$N_{bin,i}$ (top panel), which generally coincides with a 
flat rollover at the lower end of the differential size
distribution $N(x)=N_{bin,i}/\Delta x$ (middle panel),
or in the cumulative size distribution $N_{cum}(>x)$ (bottom 
panel).  The distributions shown here refer to a data set
with 10,665 solar flare events described in Section 4.1.}
\end{figure}

\begin{figure}
\plotone{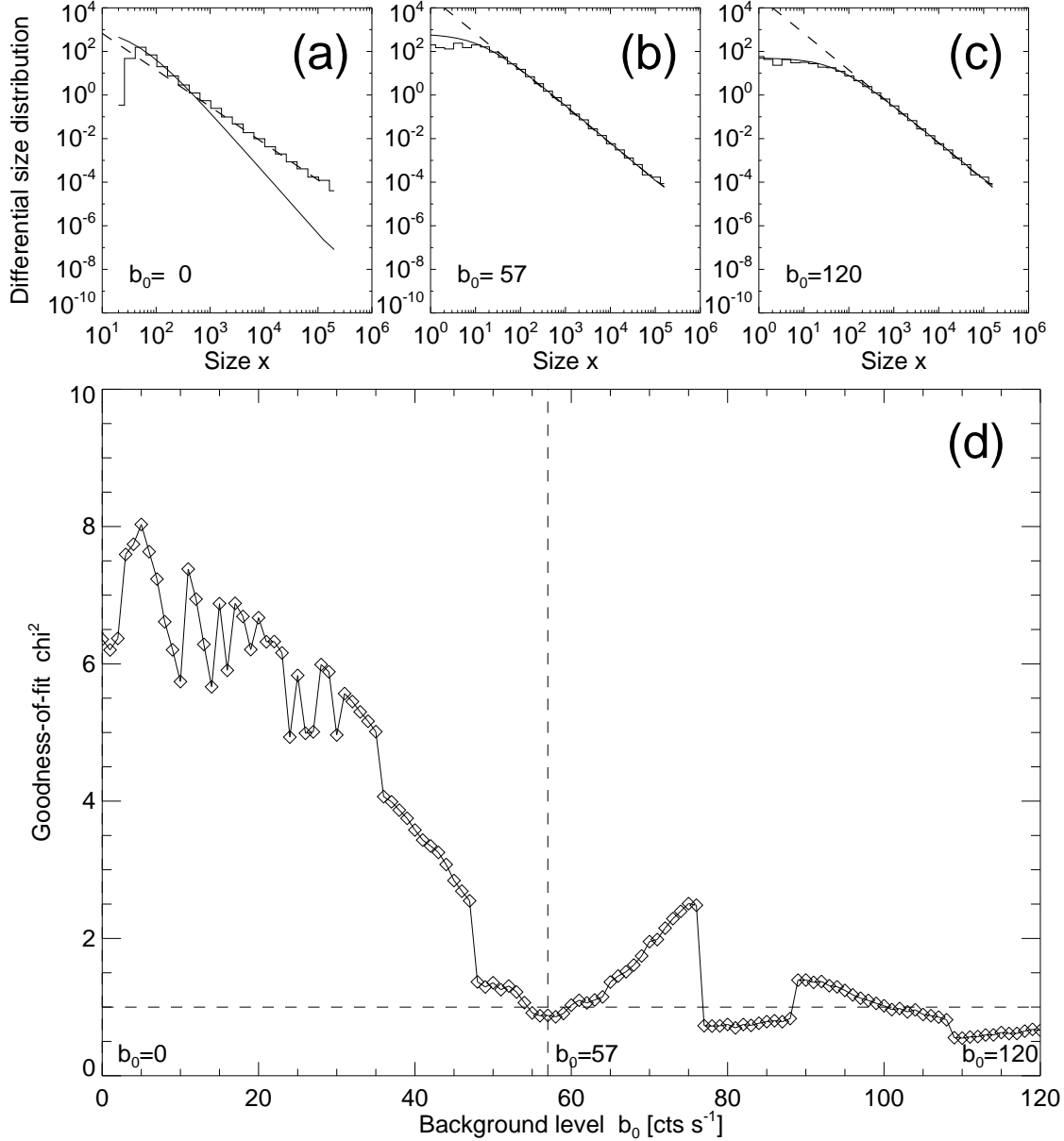}
\caption{The background subtraction procedure is shown for the
same solar flare data set as shown in Fig.~6. Three examples of
differential size distributions sampled with different background
subtractions of $b_0=0, 57, 120$ cts s$^{-1}$ 
are shown (histograms in top panels),
along with fits of thresholded power law distributions (smooth
curves in top panels) and the power law slopes (dashed lines in top
panels). The goodness-of-fit $\chi^2$ is shown as a function of 
120 different background levels ($b_0=1, 2, ...., 120$ cts s$^{-1}$) 
in the bottom panel. Note that a best-fit value of 
$\chi^2 \approx 1.0$ is obtained empirically for an optimum 
background value of $b_0=57$ cts s$^{-1}$.} 
\end{figure}

\begin{figure}
\plotone{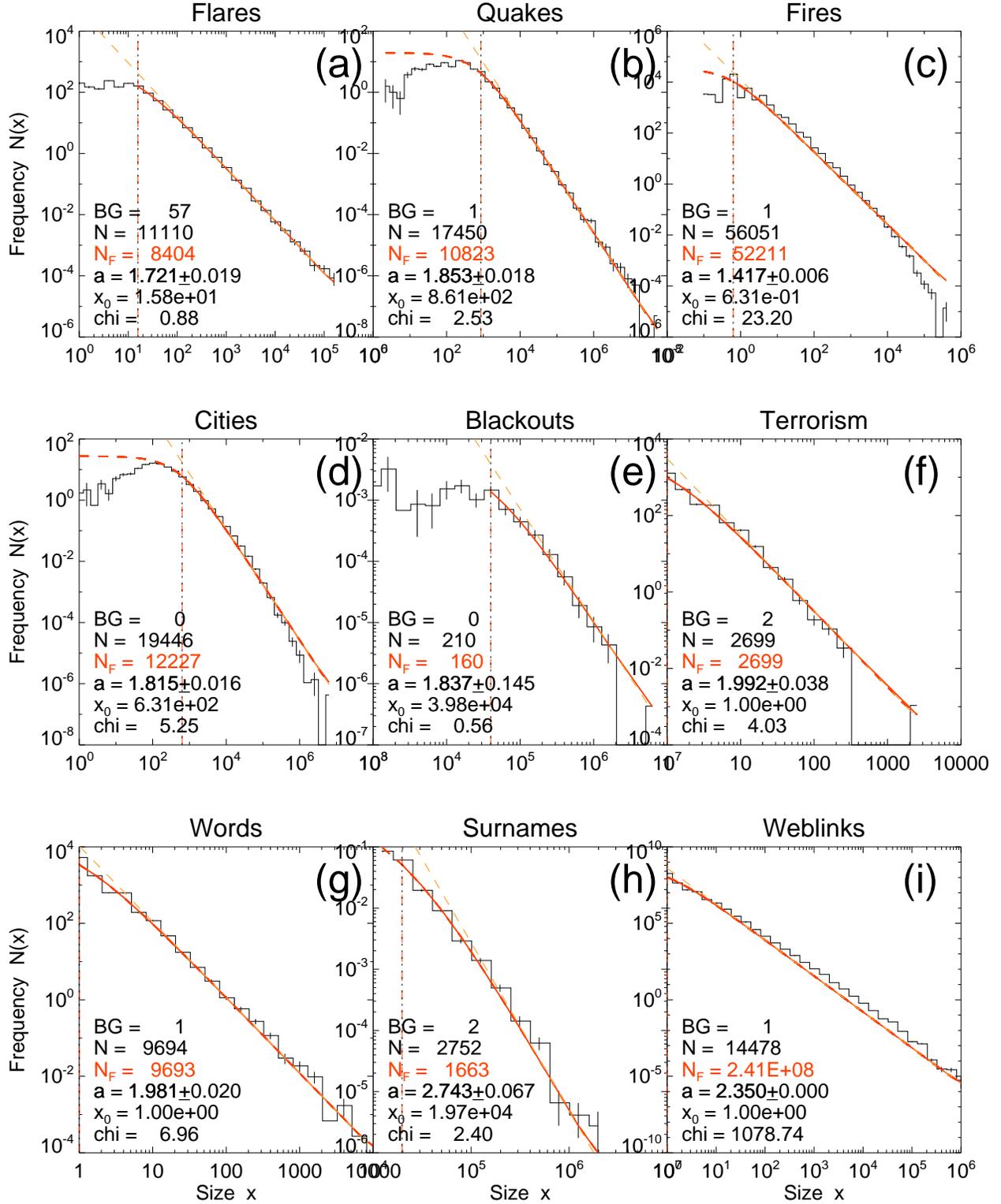}
\caption{Differential size distributions (histogram) with fits
(red curve) of thresholded power laws for 9 empirical data sets 
from Clauset et al.~2009. A straight power law is indicated with
a dashed orange line. The threshold $x_0$ found from the fit is
indicated with a vertical dotted line. 
The number of all events $(N)$, of the partial events in the
range of $x > x_0$, the power law slopes $a$, the backgrounds $BG$,
the thresholds $x_0$, and the goodness-of-fit $\chi^2$ are listed 
also.}
\end{figure}

\begin{figure}
\plotone{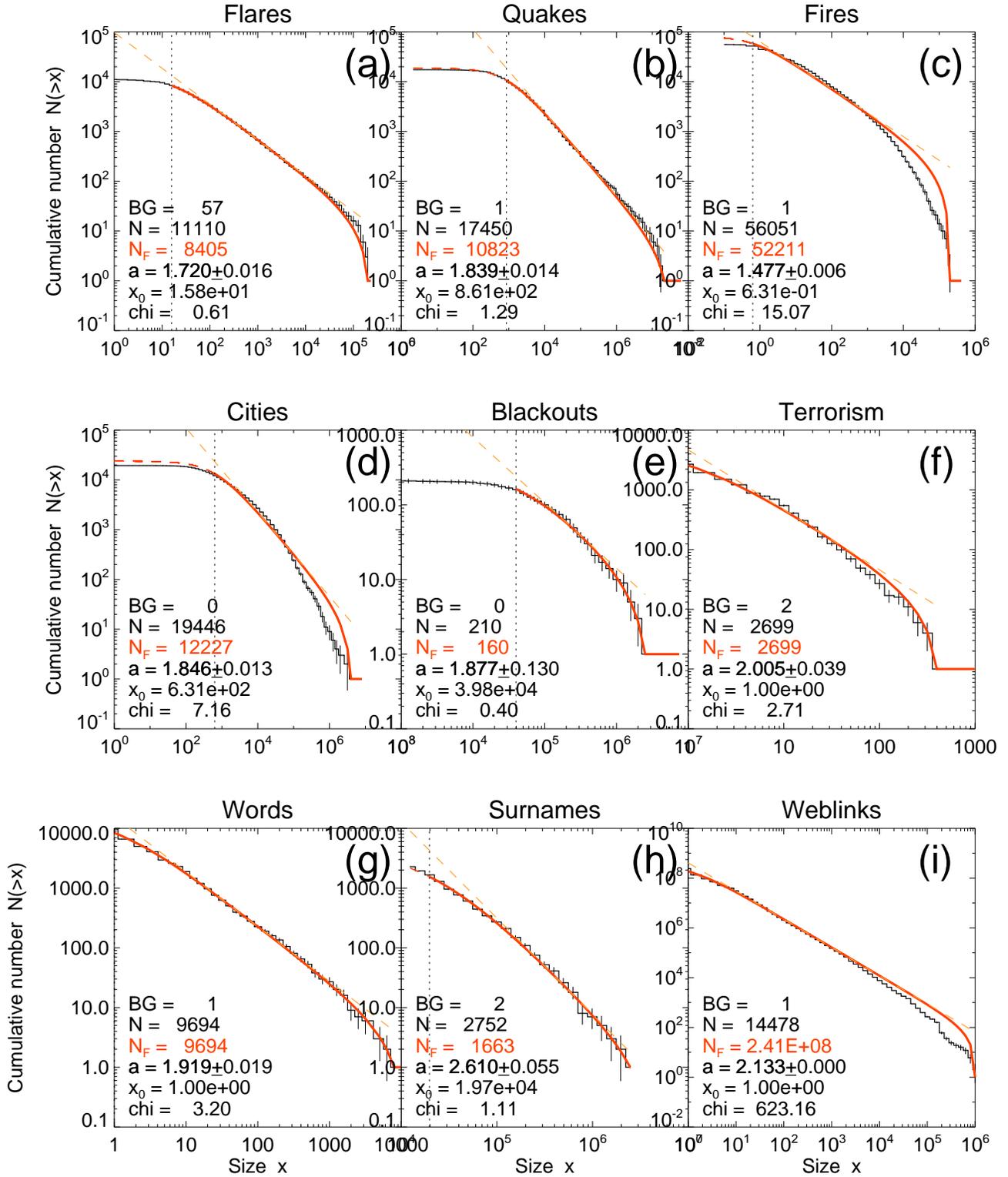}
\caption{Cumulative size distributions (histogram) with fits
(red curve) of thresholded power laws for 9 empirical data sets 
from Clauset et al.~2009. Representation similar to Figure 8.}
\end{figure}

\begin{figure}
\plotone{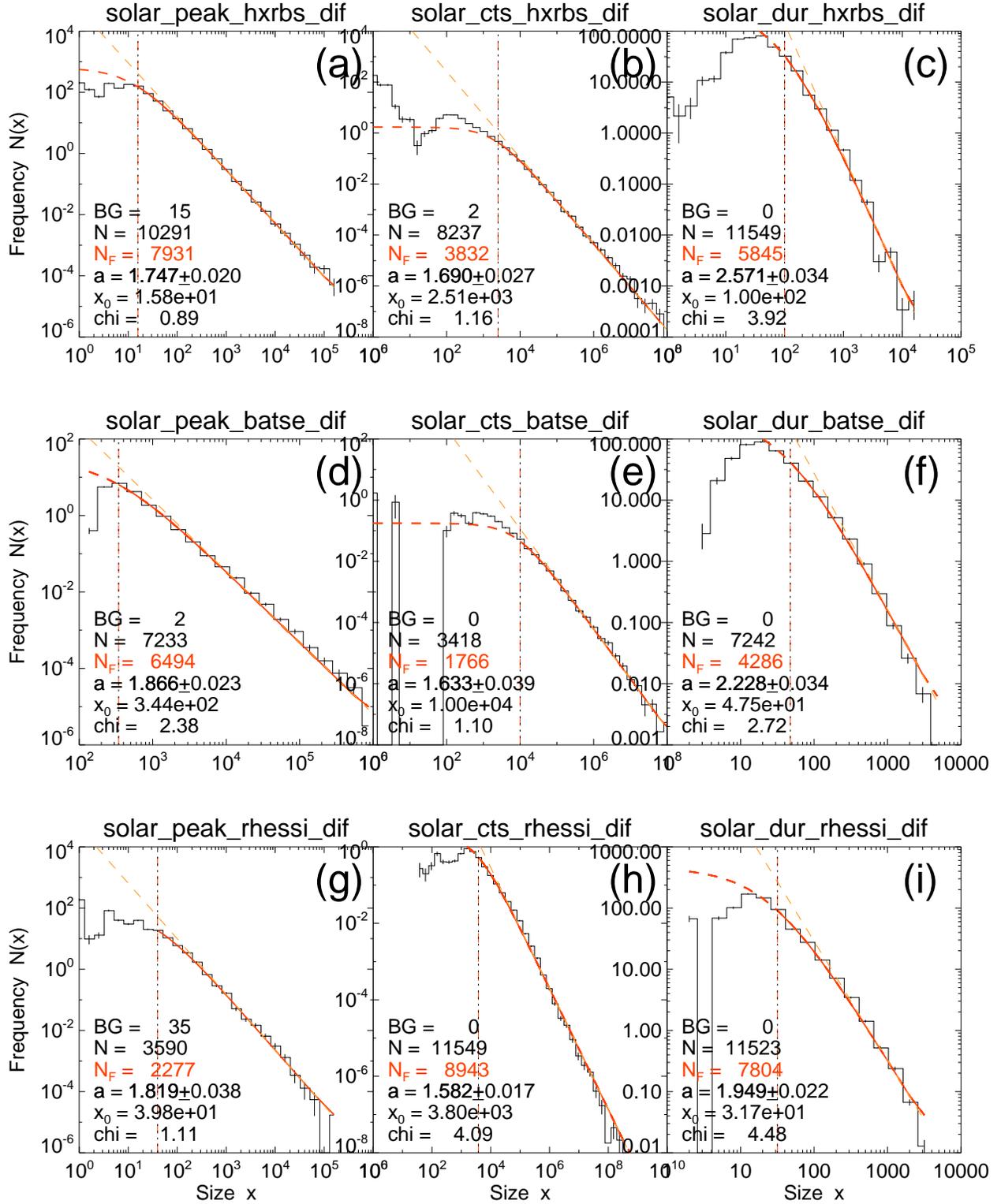}
\caption{The differential size distributions of hard X-ray peak counts (left), 
total counts (middle), and durations of solar flare events (right panel),
are shown from HXRBS/SMM (top row), BATSE/CGRO (middle row),
and RHESSI data (bottom row).}
\end{figure}

\begin{figure}
\plotone{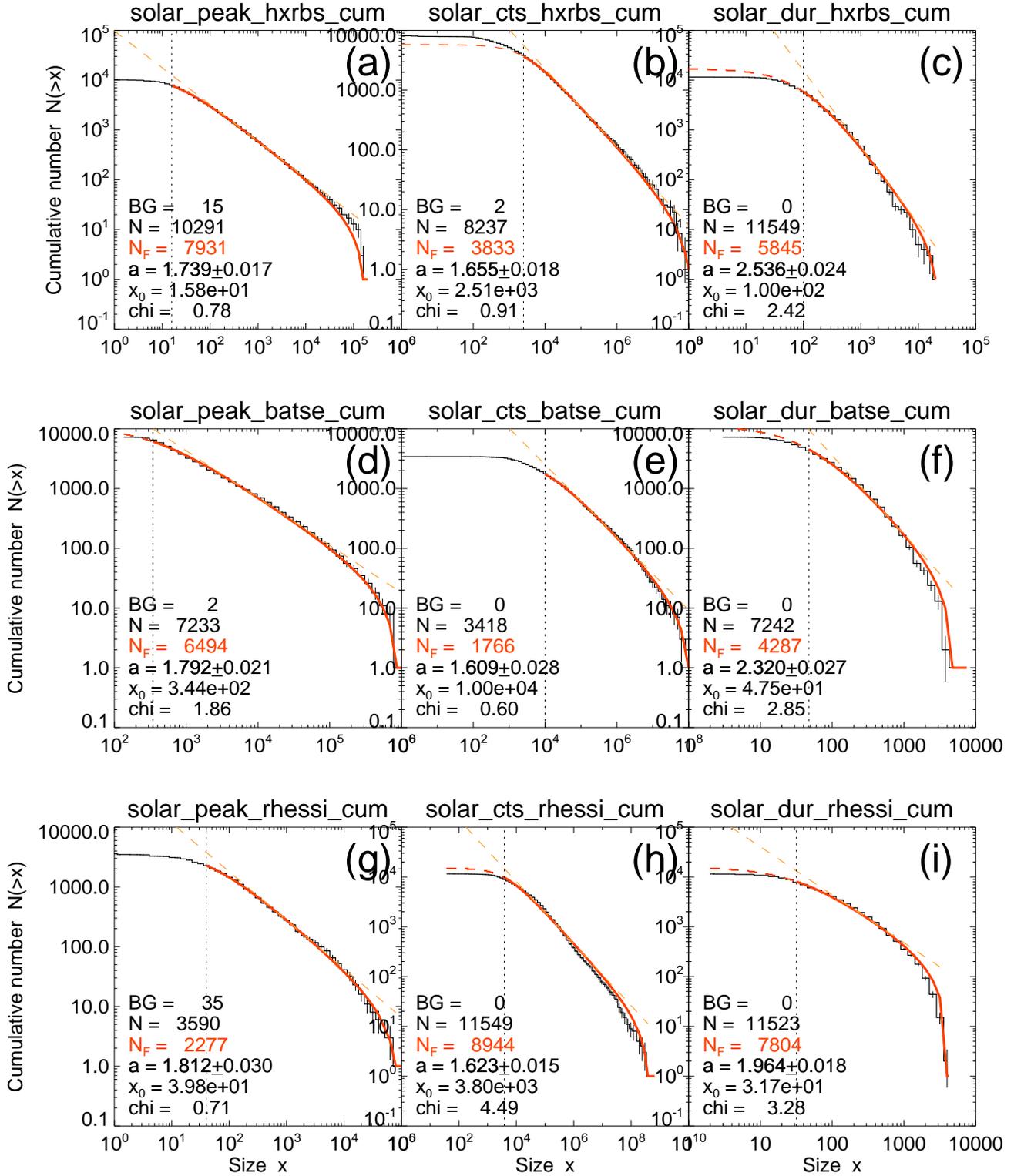}
\caption{The cumulative size distributions of hard X-ray peak counts (left), 
total counts (middle), and durations of solar flare events (right panel),
are shown from HXRBS/SMM (top row), BATSE/CGRO (middle row),
and RHESSI data (bottom row). Representation similar to Fig.~10.}
\end{figure}

\begin{figure}
\plotone{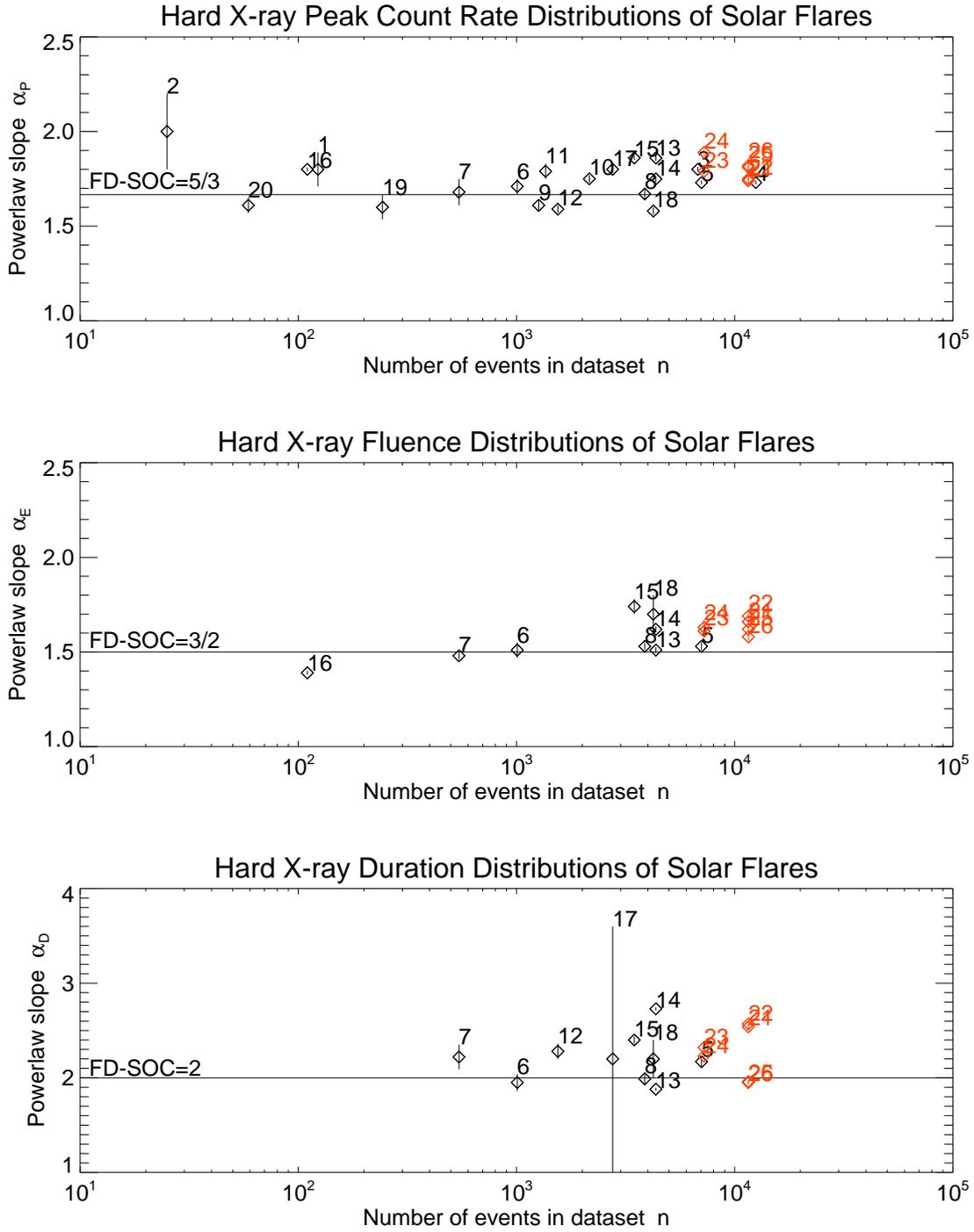}
\caption{The power law slopes of hard X-ray peak counts (top),
fluences (middle), and flare durations (bottom) are shown as a
function of the number of events $n$ in each data set published
earlier (small diamonds) or analyzed in this study (red diamonds).
The numbers refer to the references of Table 3.}
\end{figure}

\begin{figure}
\plotone{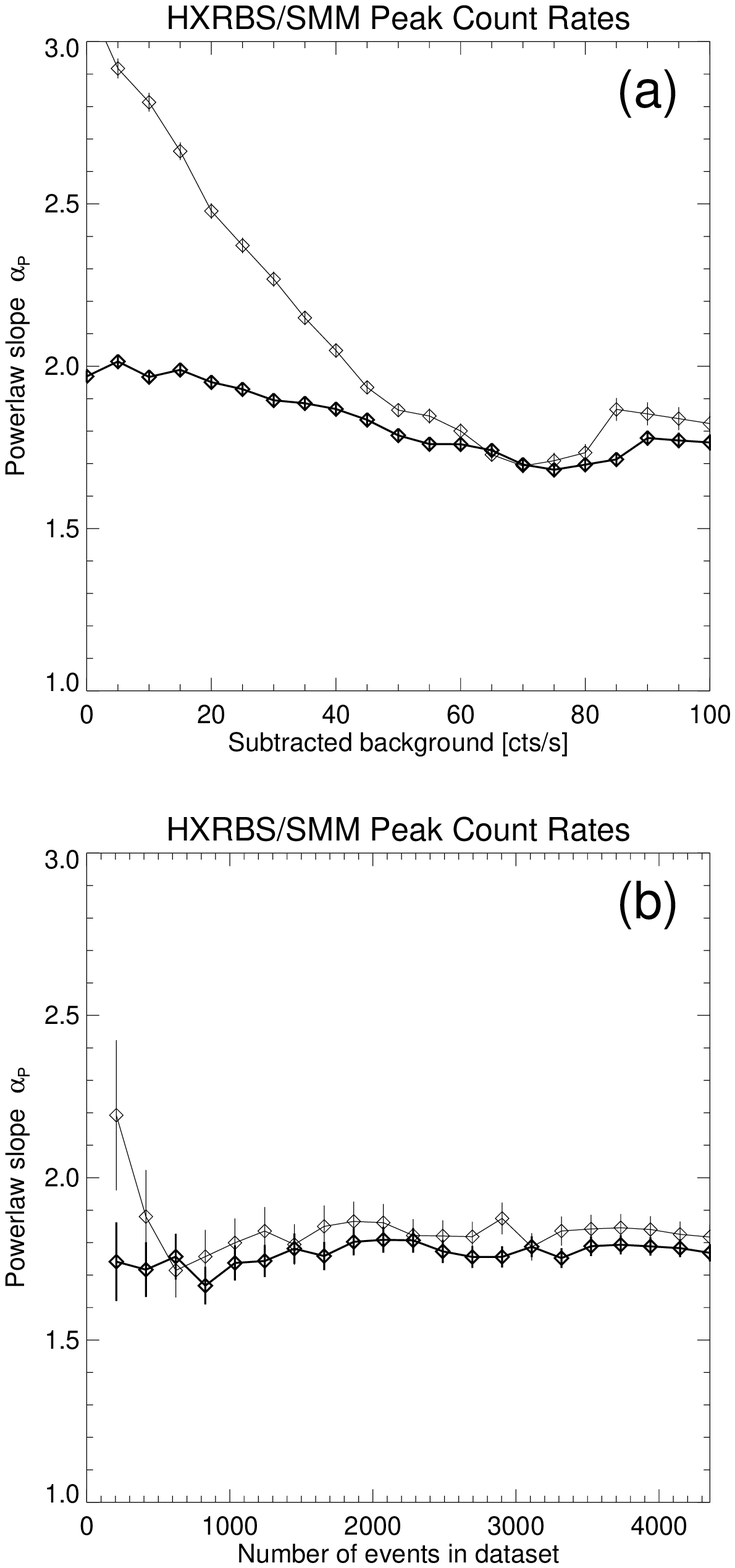}
\caption{The dependence of the power law slope on the amount
of subtracted background (a) and on the number of events
as a function of time (b) is shown for the cumulative
size distribution fits (thick linestyle) and the differential
size distribution fits (thin linestyle), for the HXRBS/SMM
$>25$ keV data set 1980-1982 with 6461 events. A background
of 57 cts/s is subtracted in the subsets in (b).}
\end{figure}

\begin{figure}
\plotone{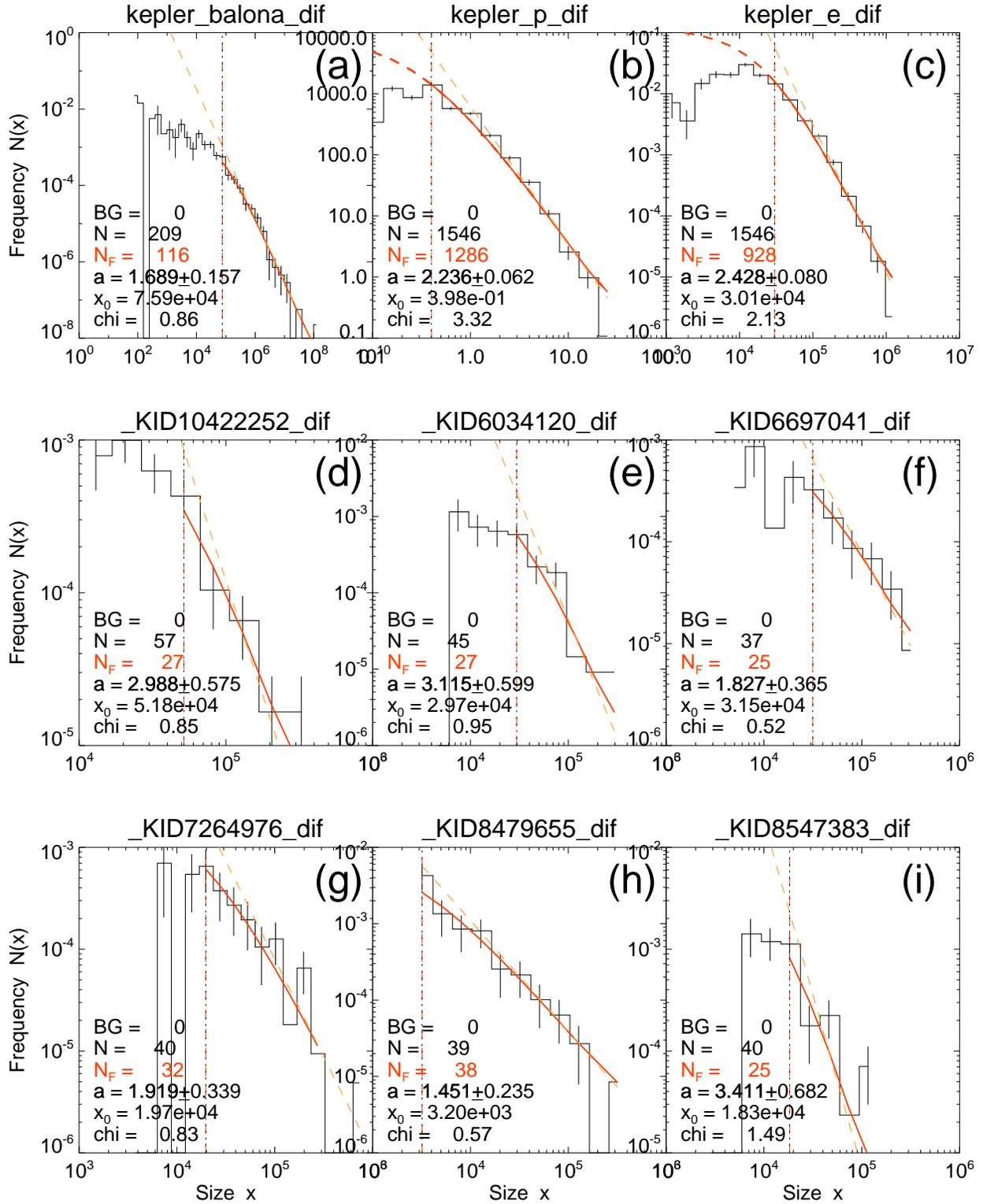}
\caption{Differential size distributions
of stellar flares observed from the Kepler mission,
described in Balona (2015) and in Shibayama et al.~(2013).}
\end{figure}

\begin{figure}
\plotone{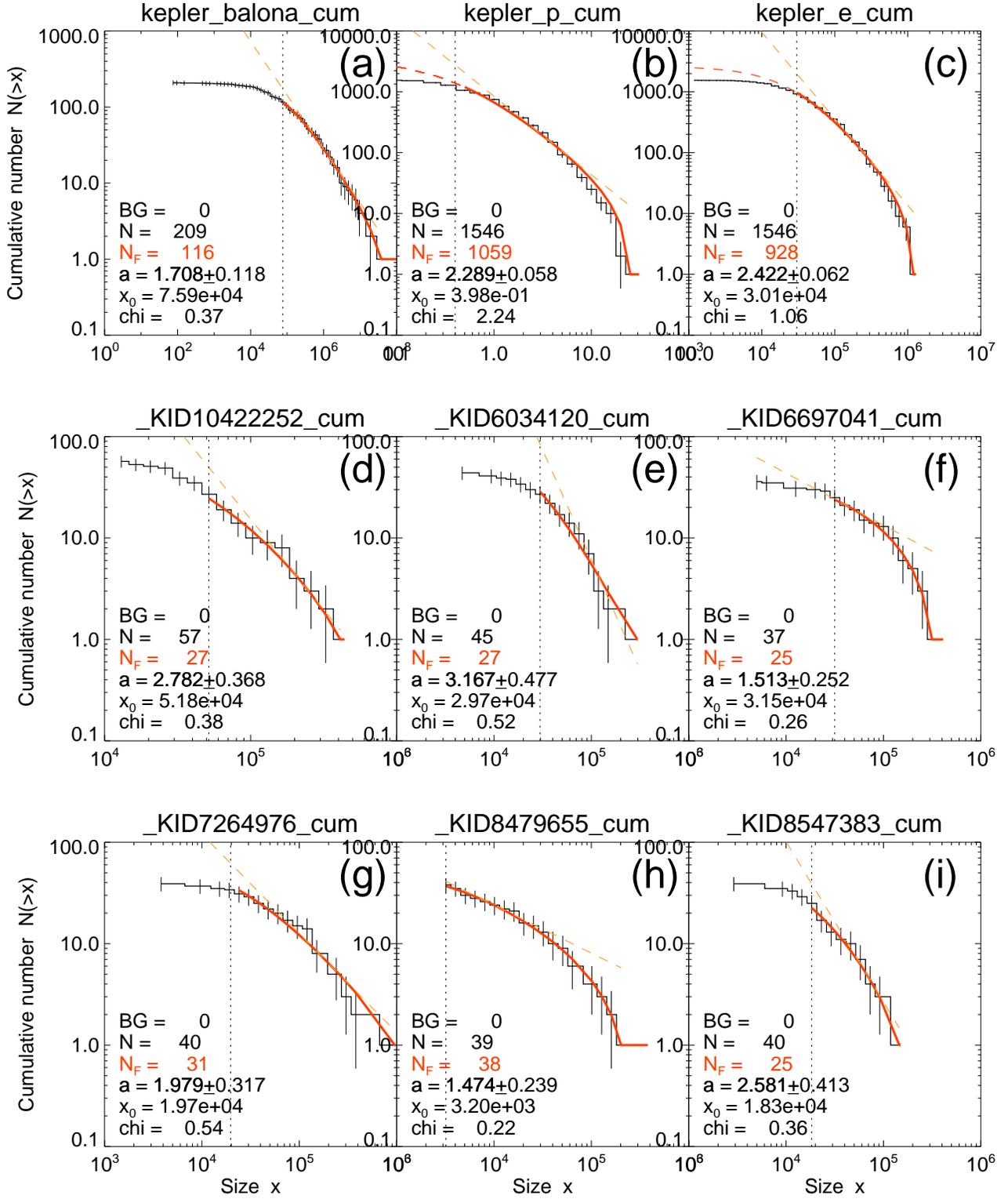}
\caption{Cumulative size distributions
of stellar flares observed from the Kepler mission,
described in Balona (2015) and in Shibayama et al.~(2013).}
\end{figure}

\begin{figure}
\plotone{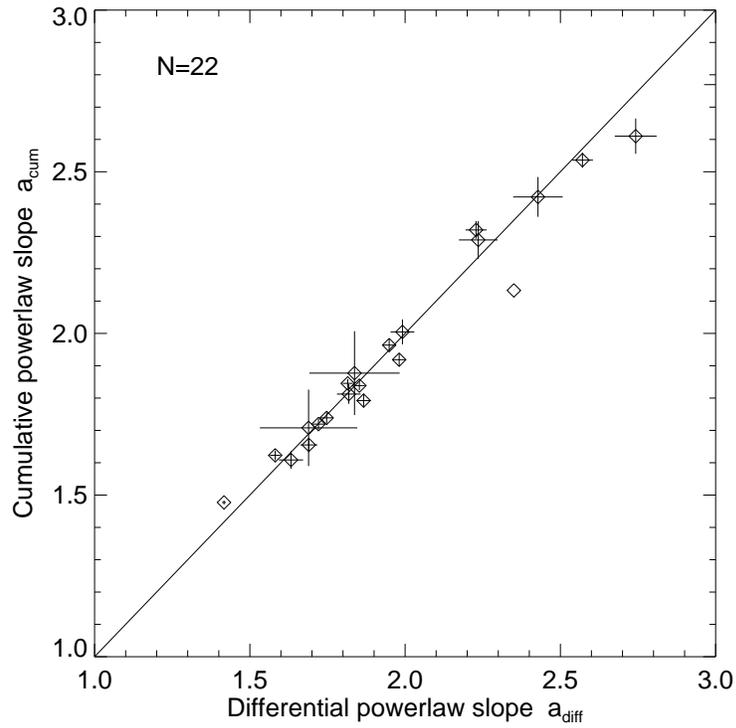}
\caption{The accuracy of the fitted power law slope 
$\alpha \pm \sigma_\alpha$
is shown as a scatterplot between the differential size distribution 
fit method (x-axis) and the cumulative size distribution method 
(y-axis), for the 18 empirical and solar data sets shown in
Figures 8$-$11.}
\end{figure}

\end{document}